\documentclass[12pt]{article}
\pdfoutput=1
\usepackage{jheppub}
\usepackage{amsmath}
\usepackage{amsfonts}
\usepackage{amssymb}
\usepackage{graphicx}

\usepackage[export]{adjustbox}
\newcommand{\bmat}{\left(\begin{array}}
\newcommand{\emat}{\end{array}\right)}

\def\yzero{\smash{\hbox{$y\kern-4pt\raise1pt\hbox{${}^\circ$}$}}}

\def\beq{\begin{equation}}
\def\eeq{\end{equation}}
\def\beqa{\begin{eqnarray}}
\def\eeqa{\end{eqnarray}}

\def\-{\hphantom{-}}
\def\ov{\overline}
\def\s2{\frac{1}{\sqrt2}}

\def\beq{\begin{equation}}
\def\eeq{\end{equation}}
\def\beqa{\begin{eqnarray}}
\def\eeqa{\end{eqnarray}}
\def\tr{{\rm tr \,}}

\def\IF{\relax{\rm I\kern-.18em F}}
\def\II{\relax{\rm I\kern-.18em I}}

\def\Dsl{\,\raise.15ex\hbox{/}\mkern-13.5mu D} 

\def\IC{{\bf{C}}}
\def\IS{{\bf {S}}}
\def\IR{{\bf {R}}}
\def\IZ{{\bf {Z}}}
\def\IX{{\bf {X}}}

\def\IP{{\bf {P}}}
\def\ID{{\bf {D}}}

\def\CN{{\cal {N}}}

\def\NN{{\cal {N}}}

\newcommand{\drawsquare}[2]{\hbox{%
\rule{#2pt}{#1pt}\hskip-#2pt
\rule{#1pt}{#2pt}\hskip-#1pt
\rule[#1pt]{#1pt}{#2pt}}\rule[#1pt]{#2pt}{#2pt}\hskip-#2pt
\rule{#2pt}{#1pt}}

\newcommand{\fund}{~\raisebox{-.5pt}{\drawsquare{6.5}{0.4}}~}
\newcommand{\antifund}{~\overline{\raisebox{-.5pt}{\drawsquare{6.5}{0.4}}}~}

\newcommand{\symm}{~\raisebox{-.5pt}{\drawsquare{6.5}{0.4}}\hskip-0.4pt%
        \raisebox{-.5pt}{\drawsquare{6.5}{0.4}}~}

\newcommand{\asymm}{~\raisebox{-3.5pt}{\drawsquare{6.5}{0.4}}\hskip-6.9pt%
        \raisebox{3pt}{\drawsquare{6.5}{0.4}}~}

\newcommand{\antiasymm}{~\overline{\raisebox{-3.5pt}{\drawsquare{6.5}{0.4}}\hskip-6.9pt%
        \raisebox{3pt}{\drawsquare{6.5}{0.4}}}~}

\catcode`\@=11   
\newdimen\@rotdimen
\newbox\@rotbox  

\def\@vspec#1{\special{ps:#1}}
\def\@rotstart#1{\@vspec{gsave currentpoint currentpoint translate
   #1 neg exch neg exch translate}}
\def\@rotfinish{\@vspec{currentpoint grestore moveto}}
\def\@rotr#1{\@rotdimen=\ht#1\advance\@rotdimen by\dp#1%
   \hbox to\@rotdimen{\hskip\ht#1\vbox to\wd#1{\@rotstart{90 rotate}%
   \box#1\vss}\hss}\@rotfinish}
\def\@rotl#1{\@rotdimen=\ht#1\advance\@rotdimen by\dp#1%
   \hbox to\@rotdimen{\vbox to\wd#1{\vskip\wd#1\@rotstart{270 rotate}%
   \box#1\vss}\hss}\@rotfinish}%
\def\@rotu#1{\@rotdimen=\ht#1\advance\@rotdimen by\dp#1%
   \hbox to\wd#1{\hskip\wd#1\vbox to\@rotdimen{\vskip\@rotdimen
   \@rotstart{-1 dup scale}\box#1\vss}\hss}\@rotfinish}%
\def\@rotf#1{\hbox to\wd#1{\hskip\wd#1\@rotstart{-1 1 scale}%
   \box#1\hss}\@rotfinish}%
\def\rotate{\@ifnextchar[{\@rotate}{\@rotate[l]}}
\def\@rotate[#1]#2{\setbox\@rotbox=\hbox{#2}\@nameuse{@rot#1}\@rotbox}

\catcode`\@=12

\begin{document}

\makeatletter
\@addtoreset{equation}{section}
\makeatother
\renewcommand{\theequation}{\thesection.\arabic{equation}}
\pagestyle{empty}
\vspace*{0in}
\rightline{IFT-UAM/CSIC-24-135}
\vspace{0.8cm}
\begin{center}
\Large{\bf End of the World Brane Dynamics \\ in Holographic  4d $\NN=4$ $SU(N)$\\ with 3d $\NN=2$ Boundary Conditions}
\\[7mm] 

\large{Jes\'us Huertas,   Angel M. Uranga \\[3mm]}
\footnotesize{Instituto de F\'{\i}sica Te\'orica IFT-UAM/CSIC,\\[-0.3em] 
C/ Nicol\'as Cabrera 13-15, 
Campus de Cantoblanco, 28049 Madrid, Spain}\\ 
\footnotesize{\href{j.huertas@csic.es}{j.huertas@csic.es},  \href{mailto:angel.uranga@csic.es}{angel.uranga@csic.es}}

\vspace*{8mm}

\small{\bf Abstract} \\
\end{center}
\begin{center}
\begin{minipage}[h]{\textwidth}
We consider 4d $\NN=4$ $SU(N)$ super Yang-Mills on half-space with boundary conditions defined by configurations of Gaiotto-Witten NS5- and D5-branes modded out by a rotated orientifold 5-plane breaking an extra half of the supersymmetries. We focus on configurations in which this O5’-brane is split by an NS5-brane into two oppositely charged halves, leading to a breaking of supersymmetry down to 3d $\NN=2$ at the local level. The 5-brane configurations turn into non-trivial $(p,q)$ webs and lead to non-trivial brane and gauge theory phenomena,  including a  5d $\IZ_2$ global gauge anomaly cancelled by a novel 6d global anomaly inflow, and the appearance of localized 3d $\NN=2$ Chern-Simons couplings and the 3d parity anomaly at the boundary of the 4d theory. The systems have explicit gravity duals, given by orientifolds of the AdS$_4\times\IS^2\times\IS^2$ fibrations over a Riemann surface dual to the Gaiotto-Witten setups. They describe solutions with an asymptotic AdS$_5\times\IS^5$ terminating at AdS$_4$ End of the World boundary configurations, whose rich dynamics reproduces the above physical properties. They also provide the SymTFT of the  4d $\NN=4$ on half-space with a new class of boundary conditions. We explore the interplay of bulk and boundary topological couplings and draw general lessons for the classification of cobordism defects of bulk quantum gravity theories in the swampland program.
\end{minipage}
\end{center}
\newpage
\setcounter{page}{1}
\pagestyle{plain}
\renewcommand{\thefootnote}{\arabic{footnote}}
\setcounter{footnote}{0}

\tableofcontents

\vspace*{1cm}

\newpage

\section{Introduction and Conclusions}
\label{sec:intro}

One of the key ideas in recent physics developments is the interplay of bulk and edge physics in systems with a boundary. Besides condensed matter physics, this plays a fundamental role in various areas such as holographic dualities, QFTs on spacetimes with boundary, the formulation of generalized symmetries via SymTFTs, and the cobordism conjecture in the swampland program.

A beautiful illustration of the interplay of bulk and boundary dynamics in all these areas has been provided by considering 4d $\NN=4$ $SU(N)$ super Yang-Mills defined on a half-space with 3d Gaiotto-Witten boundary conditions \cite{Gaiotto:2008sa,Gaiotto:2008ak}, i.e. coupled to a 3d $\NN=4$ boundary conformal field theory (BCFT$_3$) preserving half of the superconformal symmetry $OSp(4|4)$. The classification of such BCFT boundary conditions in \cite{Gaiotto:2008sa,Gaiotto:2008ak} in terms of configurations of semi-infinite D3-branes ending on NS5- and D5-branes \cite{Hanany:1996ie} allows the construction of explicit gravity duals \cite{DHoker:2007zhm,DHoker:2007hhe,Aharony:2011yc,Assel:2011xz,Bachas:2017rch,Bachas:2018zmb}. These have been exploited as top-down versions of Karch-Randall branes \cite{Karch:2000ct,Karch:2001cw} in higher-dimensional explorations of the quantum islands program for the recovery of the black hole Page curve \cite{Almheiri:2019hni,Almheiri:2019psy, Chen:2020uac,Chen:2020hmv,Geng:2020fxl,Geng:2021mic,Uhlemann:2021nhu,Demulder:2022aij}. Moreover, based on \cite{VanRaamsdonk:2021duo}, they were advocated in \cite{Huertas:2023syg} as dynamical realizations of End of the World (ETW) branes in the cobordism conjecture \cite{McNamara:2019rup} applied to AdS$_5\times\IS^5$, in the spirit of dynamical cobordisms in other contexts \cite{Buratti:2021fiv,Angius:2022aeq,Blumenhagen:2022mqw,Blumenhagen:2023abk}\footnote{For
  related ideas, see
  \cite{Dudas:2000ff,Blumenhagen:2000dc,Dudas:2002dg,Dudas:2004nd,Hellerman:2006nx,Hellerman:2006ff,Hellerman:2007fc}
  for early references, and
\cite{Basile:2018irz,Antonelli:2019nar,GarciaEtxebarria:2020xsr,Mininno:2020sdb,Basile:2020xwi,Mourad:2021qwf,Mourad:2021roa,Basile:2021mkd,Mourad:2022loy,Angius:2022mgh,Basile:2022ypo,Angius:2023xtu,Huertas:2023syg,Mourad:2023ppi,Angius:2023uqk,Delgado:2023uqk,Angius:2024zjv,Mourad:2024dur,Mourad:2024mpg,GarciaEtxebarria:2024jfv,Angius:2024pqk}
  for recent works.}. At the topological level, the study of 4d $\NN=4$ $SU(N)$ in \cite{Witten:1998wy} paved the way to the use of holographic dual pairs to formulate generalized symmetries of a system in terms of a bulk SymTFT with one more dimension \cite{Gaiotto:2014kfa,Gaiotto:2020iye,Ji:2019jhk,Apruzzi:2021nmk,Freed:2022qnc} with physical and topological boundaries. The SymTFT picture for 4d $\NN=4$ $SU(N)$ with Gaiotto-Witten boundary conditions has been recently obtained in \cite{GarciaEtxebarria:2024jfv} by extracting the topological sector of their gravity duals (see \cite{Copetti:2024onh,Cordova:2024iti,Copetti:2024dcz,Choi:2024tri,Das:2024qdx,Bhardwaj:2024igy,Heymann:2024vvf,Choi:2024wfm} for various aspects of SymTFTs for theories on spacetimes with boundaries).

It is natural to consider the generalization of the above web of relations to less (super)symmetric contexts, and in this work we focus on a particularly natural one. We consider 4d $\NN=4$ $SU(N)$ super Yang-Mills on half-space coupled to a class of 3d $\NN=2$ supersymmetric field theories (generically flowing to BCFT$_3$'s), i.e. preserving 1/4 of the supersymmetries, with a tractable gravity dual. Our models are based on performing an orientifold quotient (by an O5'-plane spanning directions rotated with respect to the D5-branes), breaking half of the supersymmetries of an underlying Gaiotto-Witten configuration of NS5- and D5-branes, so the gravity duals are essentially simple quotients of explicitly known supergravity solutions. The existence of an explicit gravity dual allows a detailed discussion of their topological sector, equivalently, of the SymTFT description of the 4d $\NN=4$ $SU(N)$ with a new set of boundary conditions, generalizing \cite{GarciaEtxebarria:2024jfv}.

Our models improve over other classes of 3d $\NN=2$ boundary conditions of rotated NS5'- and D5'-branes considered in \cite{Hashimoto:2014nwa,Hashimoto:2014vpa}, for which the gravity dual is not known. A further improvement is that our configurations break supersymmetry down to 4 supercharges {\em locally} at the location of the orientifold plane, rather than {\em globally} due to the branes  preserving different supersymmetries located at different locations.

Our sets of 3d $\NN=2$ boundary conditions display a plethora of non-trivial brane dynamics effects, including splitting of orientifold planes across NS5-branes, semi-infinite D5-branes and non-trivial $(p,q)$ 5-brane webs. These are associated to genuinely new gauge theory phenomena arising because of the reduced supersymmetry, including 5d $\IZ_2$ gauge global anomalies and their cancellation by a novel gauge global anomaly inflow, the appearance of localized 3d $\NN=2$ chiral flavours in the fundamental representation of the 4d gauge symmetry, and related phenomena such as 3d parity anomalies and 3d Chern-Simons terms. We therefore pay special attention to these topological phenomena and their apperance in the field theory, the brane realization, the gravity dual description and their SymTFT interpretation.

Our work can additionally be regarded as a first step in addressing a more general program.  Given a quantum gravity theory, the swampland cobordism predicts the existence of cobordisms to nothing, i.e. ETW branes, which define boundary configurations. But it does not say much about what are the different kinds of such ETW branes a given bulk theory can admit, or how constrained their worldvolume dynamics are. It is natural to address these questions in highly tractable setups, such the construction of new boundary configurations for AdS$_5\times\IS^5$ via holography. Our results provide a class of new ETW branes with rich dynamics, which illustrate important lessons for the general problem. On one hand (and admittedly not surprisingly) the 5d bulk topological couplings severely restrict the possible presence of anomalies on any consistent ETW 4d boundary, as required by anomaly inflow. On the other hand, we uncover explicit examples in which higher-dimensional ETW configurations with rich anomaly patterns `trivialize' upon dimensional reduction to 4d ETW boundaries, precisely to fit within the restricted pattern dictated by the 5d bulk.

Our work opens up new directions in several areas. From the viewpoint of gauge theory and brane configurations, there are several ways to continue reducing supersymmetry both of the bulk and of the boundary theories, and to consider non-conformal theories. This is a natural path that has been beaten both from the perspective of brane model building and gauge theories, and of holography, and can be traveled down in the construction of theories on spacetimes with boundary. On the other hand, from the viewpoint of the gravitational ETW branes and cobordism defects, the study of systems with reduced supersymmetry may open a window not only to the classification of boundary configurations, but also to the discussion of non-trivial dynamics of ETW branes and an improvement in its understanding in novel situations, such as time-dependent backgrounds. We hope our work helps in making progress in these and other directions.

The paper is organized as follows. In section \ref{sec:half} we review two realizations of half supersymmetric boundary conditions for the 4d $\NN=4$ $SU(N)$ theory. In section \ref{sec:d3-ns5-d5} we revisit the branes configurations of semi-infinte D3-branes ending on NS5- and D5-branes (section \ref{sec:d3-ns5-d5-hw}), their gravity duals (section \ref{sec:d3-ns5-d5-dual}) and topological and SymTFT interpretation (section \ref{sec:d3-ns5-d5-symtft}). In section \ref{sec:o5} we revisit brane configurations including additional O5-planes parallel to the D5-branes (section \ref{sec:o5-hw}), their gravity duals (section \ref{sec:o5-dual}) and topological/SymTFT interpretation (section \ref{sec:o5-symtft}). In section \ref{sec:quarter-rotated} we briefly discuss 1/4 supersymmetric boundary 5-brane configurations using rotated NS5'- and D5'-branes (section \ref{sec:rotated-ns5-d5}) or general $(p,q)$ 5-brane webs (section \ref{sec:pq5}). 

We devote section \ref{sec:quarter-o5} to our main class of examples of 1/4 supersymmetric boundaries via the introduction of rotated O5'-planes. In section \ref{sec:fork-c0} we discuss 5-brane configurations including O5'-planes, and the effects of the splitting of the latter by the NS5-branes, both for vanishing value of the type IIB axion $C_0=0$ (section \ref{sec:co0}) or for non-trivial $C_0=1/2$ (section \ref{sec:co12}), where a novel 5d global gauge anomaly inflow mechanism is uncovered. In section \ref{sec:fork-bc-parity-anomaly} we introduce D3-branes ending on the 5-brane configuration and build boundary configurations with non-trivial Chern-Simons terms and exhibiting 3d parity anomalies. Section \ref{sec:d3-ns5-d5-o5-dual} treats the gravity duals of the previous configurations and physical effects. The gravitational solutions and the O5'-plane geometry are discussed in section \ref{sec:gravity-dual-o5prime}. In section \ref{sec:puzzle} we find and solve a puzzle between the rich higher-dimensional anomaly structure of the ETW configuration and its trivialization upon reduction to a 4d effective boundary. In section \ref{sec:parity-anomaly} we discuss the gravity dual realization of boundary configurations exhibiting 3d parity anomaly. Some interesting general lessons are presented in section \ref{sec:lesson}. In section \ref{sec:argument} we emphasize the constraint that all consistent ETW boundaries of a given bulk theory must have a universal anomaly structure to be compatible with the anomaly inflow bulk topological couplings, and in section \ref{sec:further} we present realizations of AdS$_5(\times\IS^5)$ with two boundaries to illustrate these constraints and the remaining dynamical freedom. In appendix \ref{sec:4d-global-anomaly} we present a novel example of a 4d global gauge anomaly inflow analogous to the 5d one in the main text. In appendix \ref{sec:tduals} we relate the 5-brane configurations in the main text with other T-dual setups, in particular O6/D6 configurations, and D2/D6-branes at $\IS^2/\IZ_N$ orientifolds with 4 susys.

\section{Half supersymmetric boundaries for AdS$_5\times \IS^5$ / 4d $\NN=4$ $SU(N)$ }
\label{sec:half}

In this section we review the realization of 4d $\NN=4$ $SU(N)$ SYM on half-space with maximally supersymmetric boundary conditions given by a 3d $\NN=4$ BCFT$_3$ \cite{Gaiotto:2008sa,Gaiotto:2008ak}, and its gravity dual version of AdS$5\times \IS^5$ with an End of the World (ETW) configuration \cite{DHoker:2007zhm,DHoker:2007hhe,Aharony:2011yc,Assel:2011xz,Bachas:2017rch,Bachas:2018zmb} (see also \cite{Raamsdonk:2020tin,VanRaamsdonk:2021duo,Demulder:2022aij,Karch:2022rvr,Huertas:2023syg,Chaney:2024bgx} for recent applications).

\subsection{D3-branes ending on NS5- and D5-branes}
\label{sec:d3-ns5-d5}

A useful way to describe boundary conditions for 4d $\NN=4$ $SU(N)$ SYM is to consider stacks of D3-branes ending on 5-branes. In this section we quickly review such configurations, their gravity duals, and their generalized symmetry structures for the case of half supersymmetric boundary conditions. 

\subsubsection{Brane configurations}
\label{sec:d3-ns5-d5-hw}

We consider 4d $\NN=4$ $SU(N)$ SYM in 4d half-space, coupled to a 3d $\NN=4$ BCFT$_3$. As pioneered in \cite{Gaiotto:2008sa,Gaiotto:2008ak} these systems are efficiently studied by realizing them as a Hanany-Witten brane construction\cite{Hanany:1996ie} of $N$ semi-infinite D3-branes ending on a configuration of NS5- and D5-branes, as we quickly review. 

Consider a stack of $N$ D3-branes along the directions 0123 realizing 4d $\NN=4$ $SU(N)$ SYM. We can define a boundary for this theory by letting the D3-branes end on a configuration of NS5- and D5-branes localized in the direction 3, and spanning 012, and some additional directions. In order to preserve the maximum supersymmetry, the NS5-branes span the directions 012$\,$456, while the D5-branes span the directions 012$\,$789, and there can be additional D3-branes suspended among them. The invariant information of the configuration is the total number of asymptotic D3-branes $N$, and the linking numbers of the 5-branes. We define the linking number\footnote{We purposefully use two different notions of linking number for the two kinds of branes, so as to simplify expressions such as (\ref{sum-linkings}), see \cite{Aharony:2011yc} for discussion.} for a NS5-brane (resp. D5-brane) as the net number of D3-branes ending on it from the right plus the number of D5-branes to its left (resp. minus the number of NS5-branes to its right). As uncovered in \cite{Gaiotto:2008sa,Gaiotto:2008ak}, when the $x^3$ positions of the NS5- and D5-branes follow a specific ordering, determined by their linking numbers, the limit of sending their distances to zero makes the theory flow to a 3d $\NN=4$ SCFT which provides maximally supersymmetric boundary conditions for the 4d $\NN=4$ $SU(N)$ theory. We advice the reader to check the references for further details.

The 5-branes with the same linking numbers form stacks leading to enhanced non-abelian flavour symmetries.
Labeling NS5- and D5-brane stacks with indices $a$, $b$, respectively, we denote by $n_a$ the multiplicity of NS5-branes with linking number $K_a$ and by $m_b$ that of D5-branes with linking number ${\tilde L}_b$. A simple counting from the definition of linking numbers shows that they satisfy
\beqa
N=\sum_a n_aK_a + \sum m_b{\tilde L}_b\, ,
\label{sum-linkings}
\eeqa
We refrain from further discussion of the gauge theory perspective, and instead turn to the description of these ingredients in the gravity dual.

\subsubsection{Gravity duals: End of the World branes}
\label{sec:d3-ns5-d5-dual}

In this section, we will briefly review the gravitational duals of the brane configurations described in the last section. They are given by a particular class of explicit 10d supergravity backgrounds studied in \cite{DHoker:2007zhm,DHoker:2007hhe,Aharony:2011yc,Assel:2011xz,Bachas:2017rch,Bachas:2018zmb} (see also \cite{Raamsdonk:2020tin,VanRaamsdonk:2021duo,Demulder:2022aij,Karch:2022rvr,Huertas:2023syg,Chaney:2024bgx} for recent applications).

The most general supergravity solutions preserving 16 supersymmetries
and $SO(2,3)\times SO(3)\times SO(3)$ symmetry were explicitly
constructed in \cite{DHoker:2007zhm,DHoker:2007hhe}. They are
gravitational duals of Hanany-Witten brane configurations of NS5- and
D5-branes with stacks of D3-branes as in the previous section
\cite{Aharony:2011yc}. The geometries in general have a ``bagpipe''
structure \cite{Bachas:2018zmb}, with an AdS$_4\times \IX^6$ ``bag'', with $\IX^6$ a 6d compact manifold, save for a number of AdS$_5\times \IS^5$ ``pipes'' sticking out of it. We will eventually focus on the
specific case with only one AdS$_5\times \IS^5$ ending on the bag,
which provides the gravity dual of 4d $\NN=4$ $SU(N)$ SYM on a 4d
spacetime with boundary.

The general supergravity solutions have the structure of a
fibration of AdS$_4\times \IS_1^2\times \IS_2^2$ over an oriented
Riemann surface $\Sigma$. The ansatz for the 10d metric is
\[
ds^2= f_4^2 ds^2_{AdS_4}+f_1^2 ds^2_{\IS_1^2}+f_2^2 ds^2_{\IS_2^2}+4\rho^2 |dw|^2\, .
\label{ansatz}
\]
Here $f_1$, $f_2$, $f_3$, $\rho$ are functions of a complex coordinate $w$ of $\Sigma$. There are also non-trivial backgrounds for the NSNS and RR 2-forms and the RR 4-form, for which we refer the reader to the references. 

There are closed expressions for the different functions in the above metric. As an intermediate step, we define the real functions
\[
W= \partial_w h_1\partial_{\bar w}h_2+\partial_w h_2\partial_{\bar w}h_1 
\; \;,\;\; N_1= 2h_1h_2|\partial_w h_1|^2-h_1^2 W \;\; ,\;\;N_2= 2h_1h_2|\partial_w h_2|^2-h_2^2 W\, ,
\label{wnn}
\]
in terms of two functions $h_1,h_2$ to be specified below. The dilaton is given by
\[
e^{2\Phi}=\frac{N_2}{N_1}\, ,
\label{dilaton}
\]
and the functions are given by
\[
\rho^2=e^{-\frac 12 \Phi}\frac{\sqrt{N_2|W|}}{h_1h_2}\; ,\;\; f_1^2=2e^{\frac 12\Phi} h_1^2\sqrt{\frac{|W|}{N_1}}\; ,\;\; f_2^2=2e^{-\frac 12\Phi} h_2^2\sqrt{\frac{|W|}{N_2}}\; ,\; \;f_4^2=2e^{-\frac 12\Phi} \sqrt{\frac{N_2}{|W|}}\, .
\label{the-fs}
\]

In the following we focus on the solutions describing a single asymptotic AdS$_5\times\IS^5$ region, corresponding to the gravity dual of a stack of semi-infinite D3-branes ending on a configuration of 5-branes, i.e. a 4d $\NN=4$ $SU(N)$ SYM theory with a boundary on which it couples to a Gaiotto-Witten BCFT$_3$.

The Riemann surface $\Sigma$ can be taken to correspond to a quadrant $w=r e^{i\varphi}$, with $r\in (0,\infty)$ and $\varphi\in \left[\frac{\pi}{2},\pi\right]$. At each point of the quadrant we have ${\rm AdS}_4\times \IS_1^2\times \IS_2^2$, fibered such that $\IS^2_1$ shrinks to zero size over $\varphi=\pi$ (negative real axis) and $\IS^2_2$ shrinks to zero size over $\varphi=\pi/2$ (positive imaginary axis), closing off the geometry over those edges. The general solution includes a number of 5-brane sources, describing the NS5- and D5-branes, and spikes corresponding to asymptotic regions AdS$_5\times\IS^5$, which describe 4d $\CN=4$ SYM sectors on D3-branes. 

\begin{figure}[htb]
\begin{center}
\includegraphics[scale=.09]{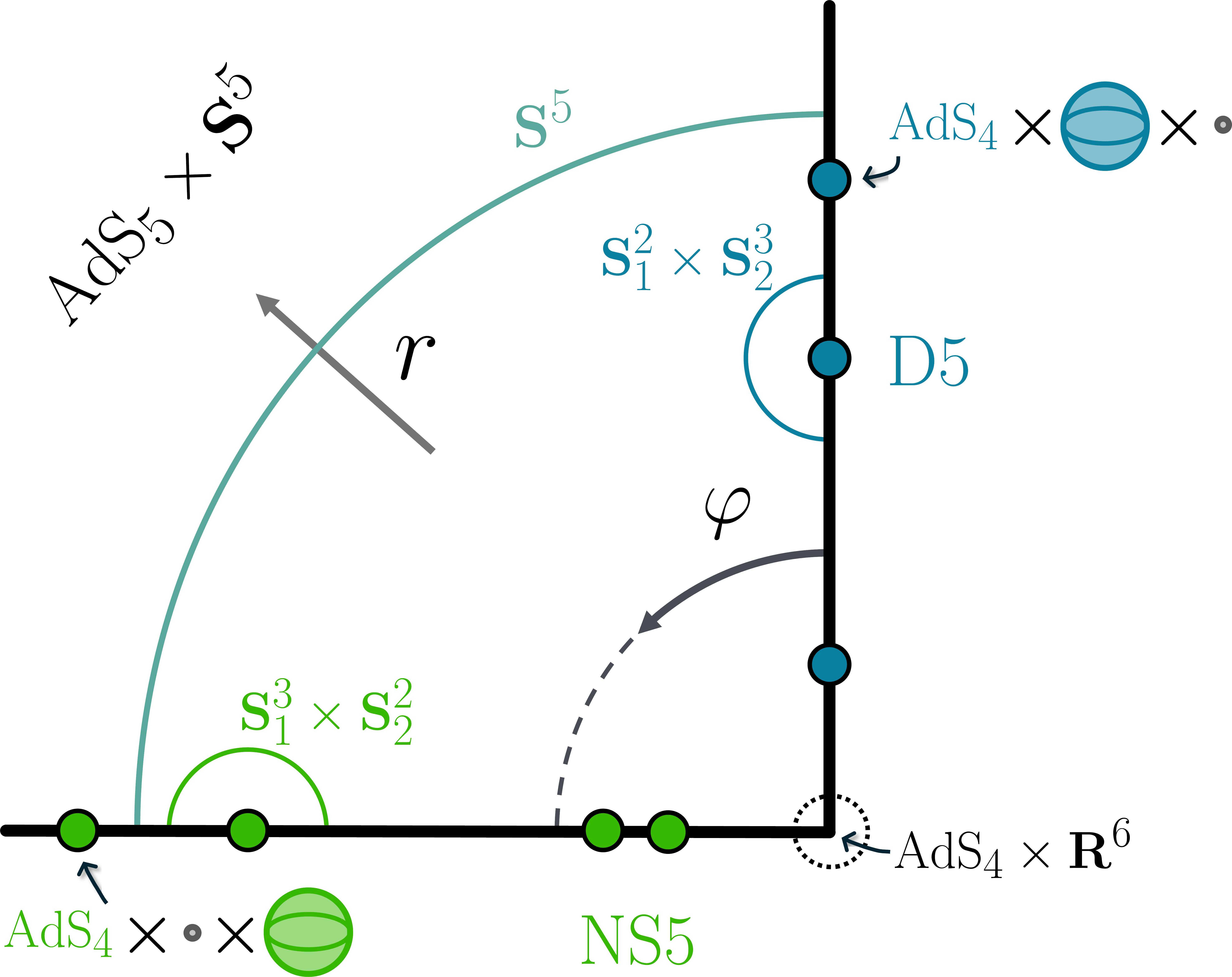}
\caption{\small Riemann surface over which the AdS$_4\times\IS^2_1\times \IS^2_2$ is fibered. The horizontal (resp. vertical) axis is the locus at which the $\IS_1^2$ (resp. $\IS_2^2$) shrinks. The green (resp. blue) dots in the horizontal (resp. vertical) axis correspond to the location of NS5-branes (resp. D5-branes). We have indicated the segments describing $\IS^5$ (purple arc) and $\IS^2\times \IS^3$'s (blue and dark green arcs) in the full fibration.}
\label{fig:quadrant}
\end{center}
\end{figure}

All these features are encoded in the quantitative expression for the functions $h_1$, $h_2$, which for this class of solutions have the structure 
\begin{subequations}
\begin{align}
h_1&= 4{\rm Im}(w)+2\sum_{b=1}^m{\tilde d}_b \log\left(\frac{|w+il_b|^2}{|w-il_b|^2}\right)\, ,\\
h_2&=-4{\rm Re}(w)-2\sum_{a=1}^n d_a \log\left(\frac{|w+k_a|^2}{|w-k_a|^2}\right)\, .
\end{align}
\label{the-hs}
\end{subequations}
In particular, the 5-brane sources are as follows: The NS5-branes are along the $\varphi=\pi$ axis, with stacks of multiplicity $n_a$ at positions $w=-k_a$, and the D5-branes are along the $\varphi=\pi/2$ axis, with stacks of multiplicity $m_b$ at positions $w=il_b$. The multiplicities of 5-branes are related to the parameters $d_a$,
${\tilde d}_b$ via
\[
n_a= 32\pi^2 d_a\in\IZ \quad ,\quad m_b= 32 \pi^2 {\tilde d}_b\in\IZ\, .
\label{quant1}
\]
The supergravity solution near one of this punctures is locally of the form of a NS5-brane spanning AdS$_4\times \IS^2_2$ (respectively a D5-brane spanning AdS$_4\times\IS^2_1$), with $n_a$ units of NSNS 3-form flux (respectively $m_b$ units of RR 3-form flux) on the $\IS^3$ surrounding the 5-brane source. The latter is easily visualised, by simply taking a segment in $\Sigma$ forming a half-circle around the 5-brane puncture, and fibering over it the $\IS^2$ fibre shrinking at the endpoints, so that we get a topological $\IS^3$, see Figure \ref{fig:quadrant}. Hence, near each 5-brane puncture, the metric factorises as AdS$_4\times \IS^2\times \IS^3$ fibered along a local radial coordinate $\tilde{r}$ parametrising the distance to the puncture. The $SU(n_a)$, $SU(m_b)$ gauge symmetry on the 5-branes is the holographic dual of the enhanced non-abelian flavour symmetries (or their duals) for the BCFT$_3$ mentioned in the previous section.

Asymptotically, at $r\to\infty$ far away from the 5-branes, the two $\IS^2$'s combine with the coordinate $\varphi$ to form a $\IS^5$, so in this limit we have AdS$_5\times\IS^5$, with $N$ units of RR 5-form flux.  The parameter $N$ will shortly be related to other quantities in the solution. The $\IS^5$ is manifest in Figure \ref{fig:quadrant}, by fibering $\IS_1^2\times\IS_2^2$ over an arc $\varphi\in[\pi/2,\pi]$ at fixed $r>|k_a|,|l_b|$, so that each of the two $\IS^2$ shrinks to zero size at one of the endpoints of the arc, leading to a topological
$\IS^5$.

The 5-form flux over the $\IS^5$ jumps when the arc crosses one of the 5-brane punctures, i.e. the flux can escape through the 5-branes. This is the gravitational manifestation that some number of D3-branes ends on the 5-brane, so the change in the 5-form flux encodes the linking number of the 5-branes in the corresponding stack. Following \cite{Aharony:2011yc}, in the $\IS^2\times\IS^3$ geometry around the stacks of $n_a$ NS5-branes and of $m_b$ D5-branes, there are non-trivial integrals
\[
\int_{\IS^2_2} C_2=K_a\quad,\quad \int_{\IS^3_1} H_3=n_a\quad ;\quad \int_{\IS_1^2} B_2={\tilde L}_b\quad,\quad \int_{\IS^3_2} F_3=m_b\, .
\label{the-fluxes}
\] 
The change in the flux 
\begin{equation}
{\tilde F_5}=F_5+B_2F_3-C_2H_3\, ,
\label{the-f5}
\end{equation}
with $F_5=dC_4$, upon crossing an NS5-brane (resp. D5-brane) is given by $n_aK_a$ (resp. $m_b{\tilde L}_b$), with no sum over indices. The 5-form flux decreases in such crossings until we reach the region $r<|k_a|,|l_b|$, in which there is no leftover flux, so that the $\IS^5$ shrinks and spacetime ends in an smooth way at $r=0$, see Figure \ref{fig:quadrant}. Hence, the asymptotic 5-form flux $N$ and the 5-brane parameters are related as in (\ref{sum-linkings})
\begin{equation}\label{N_D3}
    N=\sum_a n_a K_a+\sum_b m_b{\tilde L}_b\, ,
\end{equation} 
Finally, the positions of the 5-brane punctures $k_a$, $l_b$ are determined by the linking numbers $K_a$, ${\tilde L}_b$ via
\begin{equation}
    K_a=32\pi \left(k_a +2\sum_b{\tilde d}_b \arctan\left(\frac {k_a}{l_b}\right)\right)\quad , \quad
    {\tilde L_b}=32\pi \left(l_b +2\sum_a d_a \arctan\left(\frac {k_a}{l_b}\right)\right)\, .\quad 
    \label{positions-linking}
\end{equation}
We refer the reader to the references for a derivation of this result and other related details.

\begin{figure}[htb]
\begin{center}
\includegraphics[scale=.08]{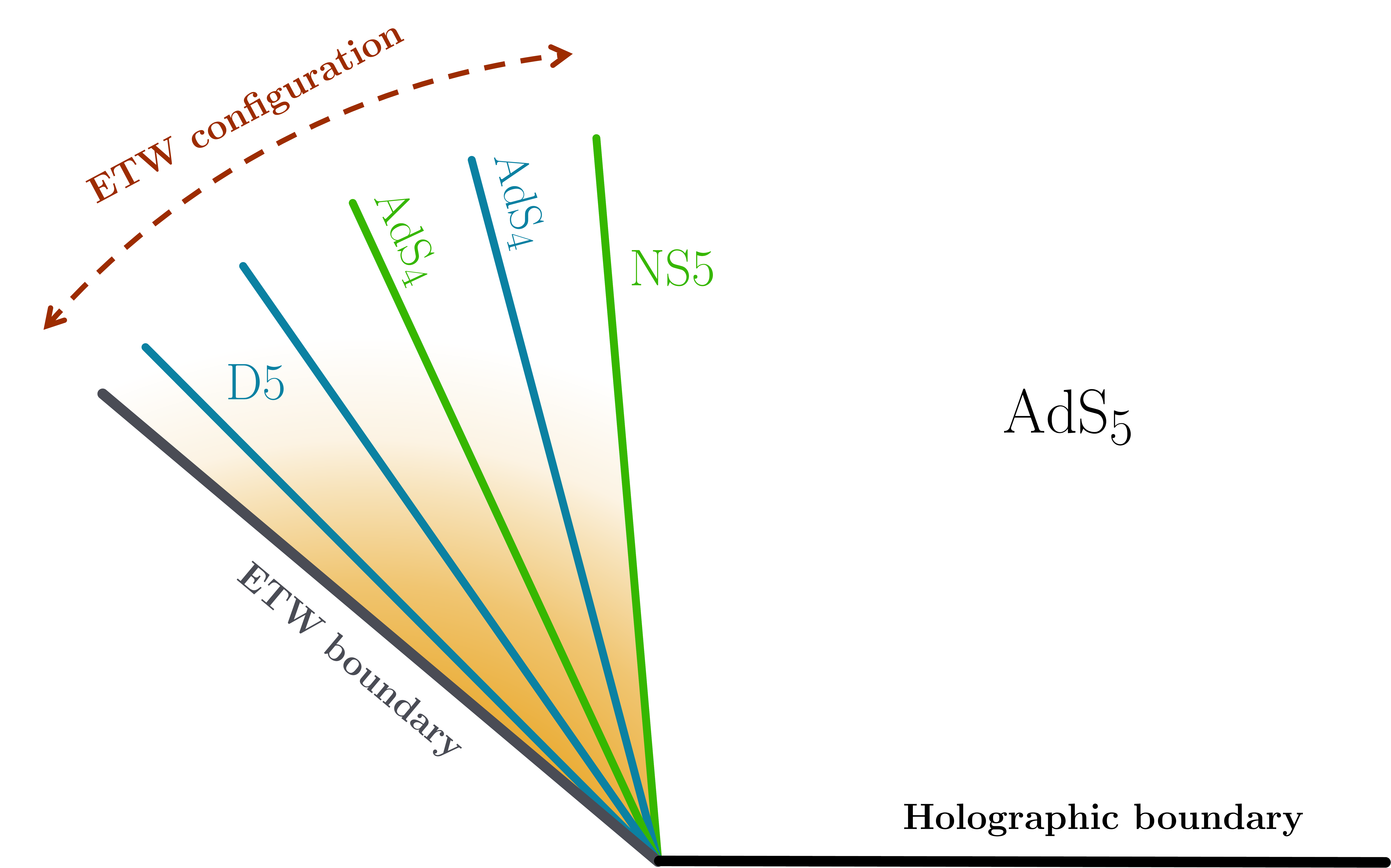}
\caption{\small The solution in Poincar\'e coordinates showin its asymptotic AdS$_5 (\times \IS^5)$ region and the ETW configuration ending spacetime.}
\label{fig:poincare}
\end{center}
\end{figure}

As emphasized in \cite{Huertas:2023syg}, the above supergravity background describes a cobordism to nothing of the asymptotic AdS$_5\times\IS^5$ via an End of the World (ETW) configuration \cite{VanRaamsdonk:2021duo}. This is shown in Figure \ref{fig:poincare} in Poincar\'e coordinates. The holographic boundary is the horizontal line representing 4d half-space. The geometry is a foliation into AdS$_4$ slices parametrized by $r=|w|$ or the Riemann surface of the solution. The asymptotic region $r\to\infty$ orresponds to the radial lines closer to the 4d holographic boundary, and is an AdS$_5$ ($\times\IS^5$) with $N$ units of 5-form flux. As one moves inward in the Riemann surface to values of $r$ corresponding to the 5-brane puncture locations, the 5d geometry hits 5-brane sources spanning AdS$_4$ slices, lines sticking out radially from the 3d boundary of the 4d holographic boundary in a fan-like structure. The 5-form flux over the $\IS^5$ decreases upon crossing the 5-branes until it is completely peeled off and spacetime ends. Hence, the solution describes and End of the World (ETW) configuration. 

We conclude with a word of caution about the above 5d picture. As is familiar, there is no scale separation \cite{Lust:2019zwm} (see  \cite{Coudarchet:2023mfs} for a review and further references) in this solution, namely the scale of the internal geometries are comparable to those of the corresponding AdS geometries, both for the asymptotic AdS$_5\times\IS^5$ and the AdS$_4\times\IX_6$ ``bag'' dual of the BCFT$_3$. As usual in holography, we may still use the lower dimensional perspective by removing the internal space via a dimensional reduction, see \cite{Huertas:2023syg}.

\subsubsection{SymTFT and ETW brane worldvolume physics}
\label{sec:d3-ns5-d5-symtft}

In this section we quickly review the structure of the topological sector of the above supergravity solutions, from the viewpoint of the 5d theory after reduction on the compact space. As recently explained in \cite{GarciaEtxebarria:2024jfv}  the 5d (quasi)topological theory is essentially the SymTFT\footnote{More precisely, it should be referred to as the Symmetry Theory, because it includes non-topological degrees of freedom. However, we will abuse language and dub it SymTFT.} encoding the generalized symmetries of the 4d $\NN=4$ $SU(N)$ theory on half-space with 3d $\NN=4$ BCFT$_3$ boundary conditions. We recap the key ideas, refering the reader to \cite{GarciaEtxebarria:2024jfv} for more detailed information.

The topological structure of the 5d solution was dubbed a SymTFT Fan, as pictured in Figure \ref{fig:poincare}. There is a physical boundary, which corresponds to the holographic boundary, in which the local  degrees of freedom of the 4d holographic dual theory are coupled. The 5d bulk contains a semi-infinite region, corresponding to the topological sector encoding the generalized symmetries of the 4d $\NN=4$ $SU(N)$ theory. As pioneered in \cite{Witten:1998wy}, the topological fields are the NSNS and RR 2-forms, which correspond to the background fields of the electric and magnetic 1-form symmetries of the 4d field theory. Their topological action is given by the dimensional reduction on $\IS^5$ of the 10 topological Chern-Simons term
\beqa
\int_{10d} F_5\, B_2\, dC_2\;\Longrightarrow \; N\int_{5d} B_2 dC_2\, ,
\label{bf-symtft}
\eeqa
which is a gapped $\IZ_N$ gauge theory. The semi-infinite 5d bulk is bounded by the ETW configuration, a fan-like structure of wedge regions separated by 5-branes, and beyond which spacetime ends. The topological fields in the wedge regions are still $B_2$, $C_2$, with an action with the structure (\ref{bf-symtft}), but with coefficient given by the local value of the RR 5-form flux, which jumps across the 5-brane domain walls. The 5-branes support non-topological degrees of freedom, corresponding to the $U(n)$ flavour symmetries of the BCFT$_3$, which can be regarded as the retraction of a 5-brane SymTFT onto a junction theory of a SymTree \cite{Baume:2023kkf}. Finally, the 5d semi-infinite bulk and the 5d wedge regions have an asymptotic infinity in the holographic direction, which at the topological level can be regarded as a topological boundary, in which topological boundary conditions are fixed for the topological 2-form fields.

At the topological level it is possible to collapse the SymTFT Fan structure (namely the different 5-branes, their intermediate wedge regions, and the ETW boundary) into a single boundary. This leads to a model-independent picture of the ETW, in the sense that it is independent of the details of the 5-brane composition. In particular,  on this effective ETW brane, there is a topological worldvolume coupling arising from the edge contribution of the  5d bulk coupling (\ref{bf-symtft}) given by
\beqa
S_{4d,{\rm top}}=N\int_{4d} B_2\,C_2\, .
\label{bf-boundary}
\eeqa
From the perspective of the 4d ETW boundary, this arises from the pile up of the individual contributions of topological couplings on the 5-branes, so its microscopic interpretation involves a model-dependent partition of $N$. Still, it is worthwhile to mention that its collective effect is to reproduce the same coupling that one picks up in the 4d holographic boundary. In the latter, this action is interpreted as a counterterm involving the background fields for the electric and magnetic 1-form symmetries, ultimately related to mixed anomalies of 1-form symmetries (see \cite{Bergman:2022otk} for a detailed discussion). 

This match may seem trivial, because it simply follows from the fact that both the 4d ETW boundary and the 4d holographic boundary are boundaries of the same 5d bulk topological theory, so they inherit the same edge topological coupling. On the other hand, it is the topological avatar of a non-trivial relation  extending that in \cite{Gubser:1999vj}; namely, that the dynamical theory on the 4d ETW boundary is given by the dynamical CFT on the 4d holographic boundary, coupled to gravity. This idea permeates the recent discussions about quantum islands \cite{Almheiri:2019hni,Almheiri:2019psy, Chen:2020uac,Chen:2020hmv,Geng:2020fxl,Geng:2021mic,Uhlemann:2021nhu,Demulder:2022aij}, where the ETW boundary supports gravitational black holes solution which radiates into a non-gravitating thermal bath described by the holographic boundary.

The interplay between topological couplings in the bulk and the different boundaries will be a useful tool in our discussion in coming sections.

\subsection{Adding O5-planes}
\label{sec:o5}

In this section we discuss a generalization of the above boundary conditions using brane configurations by allowing for the introduction of orientifold planes\footnote{It is also possible to discuss orbifold 5-planes, but they are related by S-duality to O5-planes \cite{Sen:1996na}; in particular the perturbative one is S-dual to the O5$^-$ with 2 D5-branes on top. Because of this, we will focus our discussion on the introduction of O5-planes, with additional NS5- and D5-branes, and will not consider orbifold 5-planes any further.}, as recently described in \cite{GarciaEtxebarria:2024jfv}. We will focus on orientifolds which are localised at the boundary in the direction 3 (i.e. not O3-planes along the D3-branes of the kind in \cite{Witten:1998xy}, see \cite{Hatsuda:2024lcc} for a recent discussion). We review the setup of orientifold 5-planes along 012 789, i.e. parallel to the D5-branes, so that they preserve the same symmetries and supersymmetries preserved by D5- and NS5-brane boundary conditions, so as to stay in the set of boundary conditions studied in \cite{Gaiotto:2008sa, Gaiotto:2008ak}. 

Orientifold 5-planes (O5-planes for short) are defined by quotienting by the orientifold action $\Omega R$, where $\Omega$ is worldsheet parity and $R$ is a geometric $\IZ_2$ involution acting as a coordinate flip in the directions 3456, $(x^3,x^4,x^5,x^6)\to (-x^3,-x^4,-x^5,-x^6)$.  The 6d plane fixed under $R$ is an O5-plane, and there are several discrete choices for its properties (in analogy with O3-planes in \cite{Witten:1998xy}), classified by discrete RR and NSNS backgrounds on the $\IR\IP_3$ geometry around it \cite{Hanany:2000fq}. The
O5$^-$-plane and O5$^+$-plane differ in the value of the NSNS 2-form on an $\IR\IP_2\subset \IR\IP_3$, carry RR charge $\mp 2$ (as measured in D5-units in the double cover), and project the symmetry $U(n)$ on
a stack of $n$ D5-branes on top of them down to $SO(n)$ or $USp(n)$, respectively.

There are variants of these, dubbed ${\widetilde{\rm O5}}^\pm$, which arise when the non-trivial $\IZ_2$ RR background $C_0=1/2$ is turned on at the O5-plane location. The ${\widetilde{\rm O5}}^-$ has RR charge $-1$
and can be described as the O5$^-$ with one stuck D5-brane on top, so it leads to a gauge symmetry $SO(n+1)$ when $n$ additional D5-branes are located on top of it. The ${\widetilde{\rm O5}}^+$ is an exotic
version of the O5$^+$-plane, has RR charge $+2$ and also leads to a group $USp(n)$ when $n$ D5-branes are located on top of it.

In order to describe Hanany-Witten brane configurations for systems of D3-branes ending on a set of NS5- and D5-branes in the presence of O5-planes, it is useful to consider the configuration in the covering space. There we have a $\IZ_2$ invariant system, with one stack of $N$ semi-infinite D3-branes in $x^3\to \infty$ and its image stack in $x^3\to -\infty$, ending on a ($\IZ_2$-invariant) `middle' configuration of NS5-, D5-branes, and D3-branes suspended among them, with the O5-plane sitting at $x^3=0$. The configuration is morally a back-to-back double copy of the kind of systems considered in previous sections, although with a generically non-zero number $n$ of D3-branes at the ETW boundary, which now corresponds to the O5-plane (and possible additional 5-branes on top of it).

Hence, most of the discussion regarding NS5- and D5-branes away from the O5-plane behave locally just like the configurations of NS5- and D5-branes in the previous sections. The only position at which the presence of the orientifold quotient is felt {\em locally} is precisely the O5-plane. Hence, we need to classify just different possible of 5-branes on top of the O5-plane, and each of them can produce a general class of 3d theories defining boundary conditions, by adding additional NS5- and D5-branes and their $\IZ_2$ images. 

\subsubsection{Brane configurations}
\label{sec:o5-hw}

{\bf No stuck NS5-branes}

The simplest possibility is that the O5-plane does not have additional NS5-branes on top. This possibility corresponds to the boundary conditions reducing the gauge symmetry considered in \cite{Gaiotto:2008sa,Gaiotto:2008ak}. Hence, we locally have a stack of $n$ D3-branes in the double cover, intersected by an O5-plane, and we can then use standard orientifold rules to read out the local breaking of the $SU(n)$ symmetry due to the orientifold projection. In this respect, recall that the orientifold projection of
the O5-plane on the D3-branes is of the opposite kind (namely $SO$ vs $USp$) as compared with the action on D5-branes, because the corresponding mixed open string sector has 4 DN directions \cite{Pradisi:1988xd,Witten:1995gx,Gimon:1996rq}.

For instance, the boundary conditions defined by an O5$^+$-plane reduce the symmetry $SU(n)$ to $SO(n)$, which corresponds to Class I in \cite{Gaiotto:2008ak}. In more detail, splitting the 4d $\CN=4$ vector multiplet in terms of a local 3d $\NN=4$ vector and a hypermultiplet in the adjoint of $SU(n)$, the $SO(n)$ part of the vector multiplet is even under the O5$^+$-plane action (and the remaining generators are odd), and the $\symm$ part (plus a singlet) of the hypermultiplet is even (while the $\asymm$ part is odd). This provides the definition of Dirichlet or Neumann boundary conditions for the different fields.

Similarly, the O5$^-$-plane boundary condition project the $SU(n)$ gauge symmetry down to $USp(n)$, which corresponds to Class II in \cite{Gaiotto:2008ak}. In more detail, in the covering space the $USp(n)$ part of the 3d $\NN=4$ vector multiplet is even (and the rest is odd), while the $\asymm$ part of the hypermultiplet is even
(and the $\symm+{\bf 1}$ is odd).

For the ${\widetilde{\rm O5}}^-$-plane, the configuration is equivalent to the O5$^-$-plane with one stuck D5-brane on top. Hence, to the above boundary condition we must add one localised half-hypermultiplet flavour (in the fundamental of the local $USp(n)$ symmetry and charged under the $\IZ_2\simeq O(1)$ on the stuck
D5-brane) arising from the D3-D5 open string sector. Finally, the ${\widetilde{\rm O5}}^+$-plane behaves similarly to the O5$^+$-plane, with minor modifications due to the non-trivial value of the RR axion $C_0=1/2$ on top.

In any of the above configurations, we may include additional D5-brane pairs on top of the O5-plane, which lead to extra localised flavours of the D3-brane gauge theory. We will not discuss these possibilities explicitly, but they are implicitly included in our analysis. We also insist on the fact that the configurations can be completed by adding general configurations of NS5- and D5-branes obeying the rules of the configurations in previous sections (and their $\IZ_2$ images), to define general classes of quiver gauge theories providing  Gaiotto-Witten BCFT$_3$ orientifold boundary conditions.

\medskip
{\bf With stuck NS5-brane}

A second possible local configuration corresponds to including one stuck NS5-brane on top of the O5-plane, as introduced in brane configurations in \cite{Landsteiner:1997ei}, see \cite{Giveon:1998sr} for review. 

In the covering space, we have two $\IZ_2$ image D3-brane stacks ending on the NS5-brane from opposite sides in $x^3$, and the O5-plane crossing through their intersection. The D3-brane gauge symmetry in the covering space is $SU(n)\times SU(n)$ and it projects down to a single $SU(n)$ after the orientifold. Hence, interestingly, in the quotient the $SU(n)$ gauge symmetry is not reduced by the boundary conditions. We again can consider the different possible kinds of O5-plane in turn, which differ in how the orientifold acts on the 3d $\CN=4$ hypermultiplet in the bifundamental $(\fund,\ov\fund)$ of the parent theory.

For the O5$^+$-plane, the bifundamental matter is projected down to a 2-index symmetric representation $\symm$ (plus a singlet) of the $SU(n)$ symmetry. Notice that this is compatible with the Higgsing $SU(n)\to SO(n)$ that corresponds to removing the NS5-brane from the configuration in the directions 789.

Similarly, for the O5$^-$-plane the bifundamental matter is projected down to a $\asymm$. For the ${\widetilde{\rm O5}}^-$ (equivalently an O5$^-$-plane with one stuck D5-brane), in addition to the $\asymm$ we get an additional full hypermultiplet flavour in the fundamental.  For the ${\widetilde{\rm O5}}^+$ we get a variant of the result of the O5$^+$, namely matter in the $\symm$ plus a singlet.

\subsubsection{Gravity duals: Orientifold ETW branes}
\label{sec:o5-dual}

We now discuss the gravitational dual of the configurations of 4d $\NN=4$ $SU(N)$ SYM on half-space with orientifold boundary conditions. The discussion of the corresponding supergravity solutions require a slight generalisation beyond those in section \ref{sec:d3-ns5-d5-dual}, as we review next, see \cite{GarciaEtxebarria:2024jfv} for details. In order to describe the configuration in the double cover, we need a configuration with two ($\IZ_2$ related) asymptotic AdS$_5\times\IS^5$ regions. Thus, the supergravity solutions are given by AdS$_4\times \IS^2_1\times\IS^2_2$ fibered over a Riemann surface $\Sigma$
conveniently described as a strip, parametrized by a coordinate $z$ with ${\rm Im}\, z\in [0,\pi/2]$, with the two points at infinity ${\rm Re }\, z\to\pm\infty$ corresponding to the two asymptotic AdS$_5\times\IS^5$ regions. There are NS5-brane punctures (resp. D5-brane punctures) in the lower (resp. upper) boundary\footnote{The quadrant in the ETW configurations in the previous sections can be described as a strip by simply taking $w=-e^{-z}$, with the point $w=0$, i.e. $z\to \infty$, corresponds to a closed off puncture, with vanishing 5-form flux, hence no actual asymptotic AdS$_5\times\IS^5$ region at that point \cite{Aharony:2011yc}.}.

The configuration is quotiented by $\Omega R$, with $R$ being inherited from its flat space avatar $R:(x^3,x^4,x^5,x^6)\to (-x^3,-x^4,-x^5,-x^6)$. It can be expressed in terms of the $\IS^5$ in the internal space, by using the embedding of the unit ball in $\IR^6$ as
\begin{equation}
(x^4)^2+(x^5)^2+(x^6)^2+(x^7)^2+(x^8)^2+(x^9)^2=1\, .
\end{equation}
The fixed point set is hence AdS$_4\times\IS_1^2$. In the AdS$_4\times\IS_1^2\times\IS_2^2$ fibration over the strip, it acts as antipodal identification on $\IS_2^2$ together with a reflection $z\to -{\ov z}$ on the complex coordinate on the strip. The fixed point set in the strip corresponds to the segment in the imaginary axis, but the only fixed point set in the whole geometry corresponds to the $\IS_1^2$ sitting at the upper endpoint of this segment (times AdS$_4$). We thus have an O5-plane on AdS$_4\times \IS_1^2$, and located at $z=i\pi$ in the strip; this is precisely the geometry required to preserve the same supersymmetries as the D5-branes.

\begin{figure}[htb]
\begin{center}
\includegraphics[scale=.1]{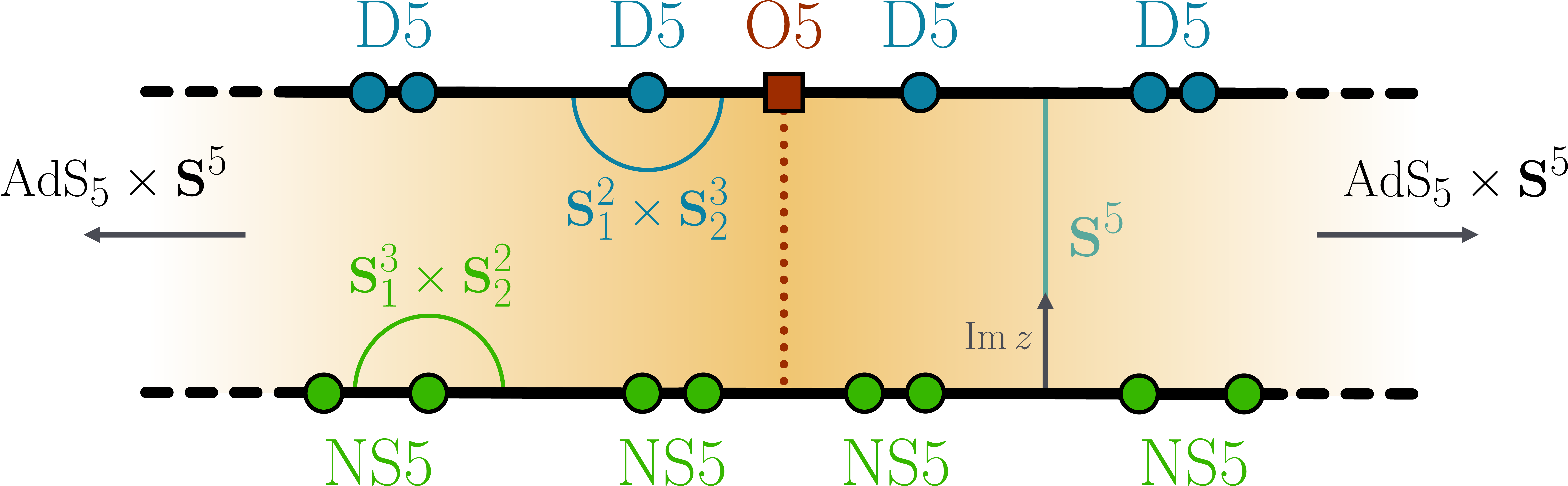}
\caption{\small Picture of the Riemann surface in the presence of and O5-plane orientifold quotient. There are two orientifold image asymptotic AdS$_5\times\IS^5$ regions, and a symmetric distribution of 5-branes. The only fixed locus in the full geometry, i.e. the O5-plane, is the AdS$_4\times\IS^2$ on top of the blue dot. For comparison with Figure \ref{fig:quadrant} we have depicted the segments corresponding to $\IS^5$ (violet vertical line) and the $\IS^2\times\IS^3$'s (blue and green arcs).}
\label{fig:orientifold}
\end{center}
\end{figure}

As explained, the positions, multiplicities and worldvolume monopole charges of NS5- and D5-branes are invariant under this $\IZ_2$ symmetry. This nicely dovetails the properties of the Hanany-Witten brane configurations for orientifold boundary conditions. From the perspective of the configuration after the quotient, we have a configuration that away from the location of the orientifold plane has the same structure as the solutions in section \ref{sec:d3-ns5-d5-dual}. The main new ingredient is thus the presence of the orientifold plane and it action on local fields and localised degrees of freedom. Hence, the ETW boundary is simply an orientifold of AdS$_5\times\IS^5$ with $n$ units of flux, whose properties encode those of the O5-plane. This was discussed in \cite{GarciaEtxebarria:2024jfv}, to which we refer the reader for details. 

\subsubsection{SymTFT and ETW brane worldvolume physics}
\label{sec:o5-symtft}

\begin{figure}[htb]
\begin{center}
\includegraphics[scale=.06]{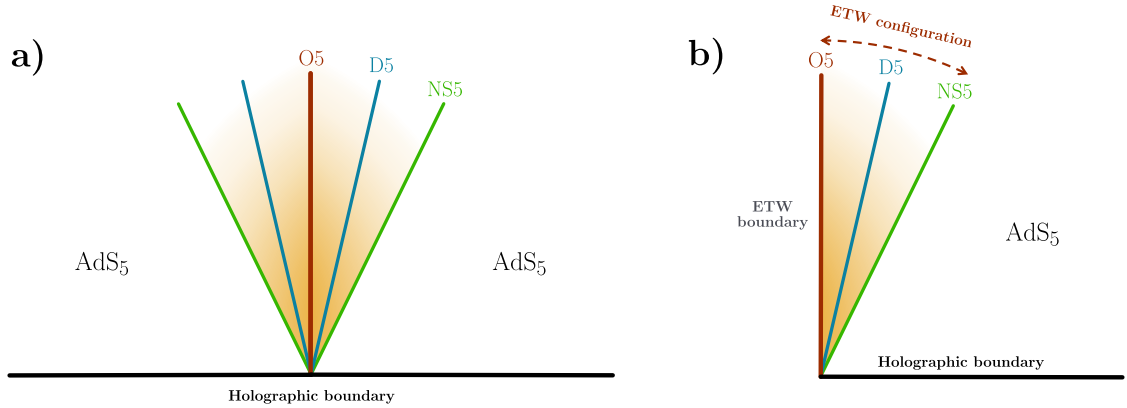}
\caption{\small a) Double cover of the SymTFT Fan of 4d $\NN=4$ $SU(N)$ SYM on half-space with orientifold boundary conditions. b) The configuration in the orientifold quotient, making manifest that the O5-plane location defines an ETW boundary.}
\label{fig:orientifold-halfsusy}
\end{center}
\end{figure}

The above 10d supergravity background can be readily translated into an effective 5d picture, whose topological sector provides a description of the SymTFT (or more precisely the Symmetry Theory) encoding the generalized symmetry structure of the 4d $\NN=4$ SYM theory on a half-space with orientifold boundary conditions. We now review the structure of this 5d theory, sketched in Figure \ref{fig:orientifold-halfsusy}, extracting the key ingredients of the Symmetry Theory for this class of models. 

In the covering space we have a full 4d Poincaré holographic boundary, with a $\IZ_2$ symmetric fan of wedges separated by 5-branes spanning AdS$_4$ slices, and the O5-plane (possibly with 5-branes on top of it) sitting in the AdS$_4$ orthogonal to the 4d holographic boundary. By folding the configuration along the vertical axis, to make the orientifold identification manifest, we reach a configuration exhibiting an ETW boundary, which corresponds to the O5-plane (possibly with 5-branes). The result away from the O5-plane is a SymTFT Fan as in section \ref{sec:d3-ns5-d5-symtft}, with the proviso that the 5-form flux does not go to zero at the location of the O5-plane. The only new ingredient is hence the presence of a new object, the O5-plane itself, whose role was discussed in \cite{GarciaEtxebarria:2024jfv}.

The discussion of topological couplings is also similar to section \ref{sec:d3-ns5-d5-symtft}. The semi-infinite AdS$_5$ region has a coupling (\ref{bf-symtft}), and we have similar structures, but with different coefficients, in the wedge regions. The only role of the O5-plane is then to possibly break some of the 5d bulk symmetries, as discussed in \cite{GarciaEtxebarria:2024jfv}. In case we collapse the whole SymTFT Fan to produce a single effective ETW boundary, its worldvolume supports the coupling (\ref{bf-boundary}). This again fits with the idea that any 4d boundary must have the same topological couplings inherited from those of the 5d bulk theory.

\section{Quarter supersymmetric boundaries via brane rotation}
\label{sec:quarter-rotated}

In this section we introduce configurations of branes providing boundary conditions for 4d $\NN=4$ $SU(N)$ preserving 4 supersymmetries. The setup is based on the well-established process of introducing the same kind of branes as in the previous section, but with suitable rotations so as to break some of the supersymmetries \cite{Elitzur:1997fh,Barbon:1997zu}, see \cite{Giveon:1998sr} for a review. In particular, we focus on the introduction of additional rotated NS5'- and D5'-branes, and O5'-planes. In the meantime, we will encounter more general  $(p,q)$ webs of 5-branes \cite{Aharony:1997ju,Aharony:1997bh}.

\subsection{Adding NS5'- and D5'-branes}
\label{sec:rotated-ns5-d5}

We start with the introduction of NS5'- and D5'-branes, which were actually considered in the context of boundary conditions for 4d $\NN=4$ $SU(N)$ SYM in \cite{Hashimoto:2014nwa,Hashimoto:2014vpa}.

We consider a stack of $N$ semi-infinite D3-branes ending on a configuration of NS5-branes along 012 456, D5-branes along 012 789, and in addition, NS5-branes along 012 678 (dubbed NS5'-branes) and D5-branes along 012 459 (dubbed D5'-branes). Note that NS5'- and D5'-branes are related to NS5- and D5-branes by a rotation of the 2-plane parametrized by 45 into the 2-plane parametrized by 89. This is an $SU(2)$ rotation, hence the complete system preserves 1/4 of the supersymmetries \cite{Berkooz:1996km}. The symmetry of the system is just $U(1)\times U(1)$ rather than $SO(3)\times SO(3)$.

The above systems were considered in \cite{Hashimoto:2014vpa,Hashimoto:2014nwa}, which studied the conditions under which they lead to superconformal BCFT$_3$ theories providing 1/4 supersymmetric boundary conditions for the 4d $\NN=4$ $SU(N)$. Interestingly enough, the new set of boundary conditions can be understood to a large extent as the combination of the boundary conditions preserving 8 supercharges associated to either the NS5/D5- and NS5'/D5'-brane systems.

The outcome that the boundary conditions are morally a superposition of those imposed by the system of NS5- and D5-branes and the system of the NS5'- and D5'-branes, should also be manifest in the SymTFT. Indeed, although the reduced supersymmetry in this case has not allowed for an explicit construction of the 10d supergravity dual of these configurations, it is easy to use the intuitions from \cite{GarciaEtxebarria:2024jfv} to guess the structure of the 5d Symmetry Theory of the 4d $\NN=4$ $SU(N)$ theory on a half-space with this kind of 3d $\NN=2$ BCFT$_3$ boundary conditions. We expect that the 5d Symmetry Theory is given by the 5d bulk SymTFT$_N$ of the 4d $\NN=4$ $SU(N)$ theory, with a SymTFT Fan of wedges corresponding to SymTFT$_{n_i}$ theories.  These wedges are separated by different kinds of 5-branes, corresponding to NS5- D5-, NS5'- or D5'-branes, and their induced D3-brane charges control the decrease of the 5-form flux (i.e. the coefficients $n_i$ of the SymTFT) as they are crossed. Eventually, the 5-form flux is completely peeled off and the system reaches a trivial SymTFT, corresponding to an $\IS^5$ compactification with no flux, so the $\IS^5$ can shrink to zero size ending spacetime, hence defining an ETW boundary.

At the topological level, we have a SymTFT Fan structure with a richer set of 5-brane theories, but in which the working of the topological 5d coupling is essentially as in section \ref{sec:half}. In particular, if we collapse the SymTFT Fan onto a single effective ETW boundary, its worldvolume supports the coupling (\ref{bf-boundary}). This again fits with the intuition that any 4d boundary, including those with lower supersymmetry such as those in this section, must have the same topological couplings inherited from those of the 5d bulk theory.

\subsection{$(p,q)$ 5-branes}
\label{sec:pq5}

Let us mention that a richer set of theories can be achieved by allowing for the introduction of $(p,q)$ 5-branes. This is interesting by itself, but will arise as a necessary ingredient in coming sections.

Let us focus on the system of NS5- and D5'-branes (a similar story holds for the NS5'- and D5-branes), and ignore the D3-branes for the time being. Recall that the NS5-brane span 012 456 while the D5'-branes span 012 459.  Hence, they share the common directions 01245, in which there is 5d Poincar\'e invariance, and they span the vertical and horizontal direction in the 2-plane 69. This is a particular instance of a more general possibility. If we introduce the complex coordinate $z=x^9+i x^6$,  a $(p,q)$ 5-brane along the direction $z=p+\tau q$ with $\tau=C_0+i/g_s$, preserves the same amount of supersymmetry as the NS5- and D5'-branes. In fact, the NS5-brane along 6 and the D5'-brane along 9 at $C_0=0$ are particular cases of $(0,1)$ and $(1,0)$ 5-branes respectively.  Thus, a natural generalization of the systems in section \ref{sec:rotated-ns5-d5} is to allow for D3-branes ending on general configuration of different stacks of $(p,q)$ 5-branes (and rotated $(p,q)$ 5'-branes, which are  $(p,q)$ versions of the NS5'- and D5-branes  similarly rotated in the directions 69, but spanning along 012 78. 

The above configurations have not been studied in the literature, but we can guess that, at the topological level, we have a 5d Symmetry Theory with a relatively straightforward SymTFT Fan generalization of the above, simply adding more general sets of 5-branes separating the SymTFT$_{n_i}$ wedges.

One can be even more general, and allow for the different $(p,q)$ 5-branes to combine into $(p,q)$ webs, considered in \cite{Aharony:1997ju,Aharony:1997bh} for the study of 5d $\NN=1$ supersymmetric gauge theories. 
When we consider such 5-brane webs as boundary configurations for D3-brane ending on them, the configurations relate to those in \cite{Aharony:1997ju}. The 5-brane configurations can be taken at the origin of their Coulomb branch, in which the web reduces to a set of semi-infinite $(p,q)$ 5-branes, with total vanishing $(p,q)$ charge,  sticking out of the origin in the $z$-plane in the directions dictated by their charges. Such systems provide superconformal boundary conditions for semi-infinite D3-branes ending at the intersection of the 5-branes. The general class can include D3-branes ending on several of these $(p,q)$ webs. Again, there is no known supergravity description for these systems (for gravity duals of general 5-brane webs, without D3-branes, see \cite{DHoker:2016ysh,DHoker:2016ujz,DHoker:2017mds,DHoker:2017zwj}), but we expect that at the topological level after reduction on the internal space we have a 5d SymTFT Fan with more involved theories separating the wedges. 

The notion of $(p,q)$ webs will turn out to be a useful language for our constructions in later sections. We will however focus on a a very particular class of them, so as to keep the system tractable enough to allow for a supergravity description. In particular they will correspond to performing an orientifold quotient to the systems in section \ref{sec:half} with an extra topping of 5-brane charges in the probe approximation.

\section{Quarter supersymmetric boundaries via O5'-planes}
\label{sec:quarter-o5}

In the previous section we have studied modifications of the brane configurations of section \ref{sec:d3-ns5-d5} which break part of the supersymmetry, via introduction of rotated branes or general $(p,q)$ 5-branes, but with no substantial modifications because the breaking to 4 susys is not implemented locally. In fact, local breaking to 4 susys requires the introduction of non-trivial $(p,q)$ 5-brane webs, which in general render the system intractable. In this section we  however describe a large class of models preserving only 4 susys already at the local level, yet remain tractable both in field theory and in the gravity dual. The configurations thus define 3d $\NN=2$ boundary conditions for 4d $\NN=4$ $SU(N)$ SYM on half space, as we have emphasized. Interestingly, the reduced supersymmetry on the ETW boundary leads to a rich dynamics, and in particular they display intricate patterns of symmetries and anomalies, with novel anomaly inflow mechanisms for e.g. global gauge anomalies. These aspects can also be  explored in the gravity dual picture, illustrating potentially interesting phenomena in the gravity dual AdS$_5\times \IS^5$ with lower supersymmetric ETW boundaries.

\subsection{5-branes with O5'-planes and the fork}
\label{sec:fork-c0}

We consider configurations of NS5- and D5-branes, as in section \ref{sec:d3-ns5-d5} and introduce O5'-planes, namely O5-planes, but, instead of parallel to D5-branes as in section \ref{sec:o5}, we take their rotated version i.e. along 012 459. Clearly, they preserve the same supersymmetries as D5'-branes. We can also possibly add NS5'- and D5'-branes, or more general $(p,q)$ 5-branes, but in order to keep the systems tractable, we mostly keep them away from the O5'-plane so that they do not interfere with our discussion below. The symmetry of the system in general is just $U(1)\times U(1)$ rather than $SO(3)\times SO(3)$.

The construction of the brane configurations is similar to section \ref{sec:o5-hw}. In the double cover, we have  a $\IZ_2$ invariant system, with one stack of $n$ semi-infinite D3-branes in $x^3\to \infty$ and its image stack in $x^3\to -\infty$, ending on a ($\IZ_2$-invariant) `middle' configuration of NS5-, D5-branes (possibly NS5'- and D5'-branes) and D3-branes suspended among them, with the O5'-plane sitting at $x^3=0$. The physics away from the O5'-plane reduces to that in previous sections, hence we need to classify just different possible of 5-branes on top of the O5'-plane. Each such configuration can produce a general class of 3d $\NN=2$ theories defining boundary conditions, by adding additional NS5-, D5-, NS5'- and D5'-branes (or more general $(p,q)$ 5-branes) and their $\IZ_2$ images, away from the O5'-plane. 

We discuss the different O5'-plane configurations, with 5-branes on top, in the following. It is convenient to consider separate discussions for the two possible choices of $C_0$ compatible with the orientifold action (under which $C_0$ is odd), namely $C_0=0$ (where the O5'-planes are the O5'$^\pm$s) and $C_0=1/2$ (where the O5'-planes are ${\widetilde{\rm O5}}^\pm$-planes).

\subsubsection{O5'-plane configurations at $C_0=0$ and forks}
\label{sec:co0}

We are interested in introducing an O5'-plane in the directions of D5'-branes, namely 012 459. In other words, we quotient under the orientifold action $\Omega {\cal R}$ with ${\cal R}:(x^3,x^6,x^7,x^8)\to (-x^3,-x^6,-x^7,-x^8)$. Since the direction 6 is flipped, but 9 is not, the configuration of 5-branes on top of the O5'-plane must be invariant under a reflection with respect to the horizontal line.

\begin{figure}[htb]
\begin{center}
\includegraphics[scale=.075]{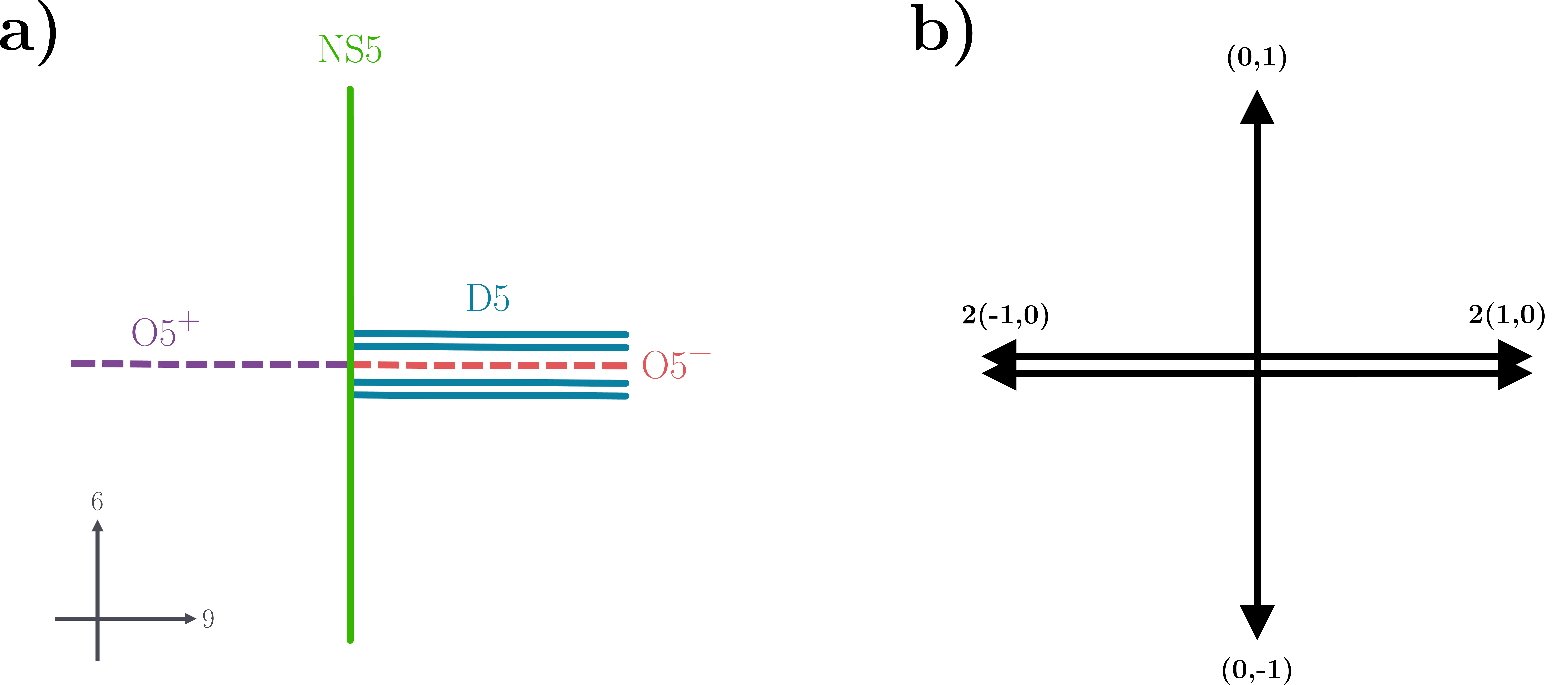}
\caption{\small The simplest fork  configuration. In (a) we draw the objects and in (b) the flow of 5-brane charges. Note that this is compatible with the orientifold action only if $C_0=0$}
\label{fig:simplest-fork}
\end{center}
\end{figure}

We focus on a relatively simple, but surprisingly rich, configuration, see Figure \ref{fig:simplest-fork}: consider a single NS5 brane along 012456 and at the origin in 378; this NS5-brane has no orientifold image, as it is mapped to itself under the O5'-plane action. In the 69 plane, the O5'-plane goes along the horizontal direction 9, while the NS5-brane goes in the vertical direction 6. However, there is a modification of this picture because of the following: the O5'-plane is split in two halves by its crossing with the NS5 in the direction 9. This implies that its RR charge flips sign in the crossing, so we have e.g. an O5'$^+$-plane for $x^9<0$ and an O5'$^-$-plane for $x^9>0$. Then, there is a mismatch of RR D5'-brane charge in the crossing, which has to be compensated in some way. The simplest is to locate 4 semi-infinite D5'-branes (as counted in the cover space) on top of the O5'$^-$-plane, see Figure \ref{fig:simplest-fork}a. The flow of 5-brane charge is in Figure \ref{fig:simplest-fork}b. We emphasize again that the configuration in these figures is compatible with the orientifold action only if $C_0=0$, and the case $C_0=1/2$ will be discussed in section \ref{sec:co12}. 

\begin{figure}[htb]
\begin{center}
\includegraphics[scale=.075]{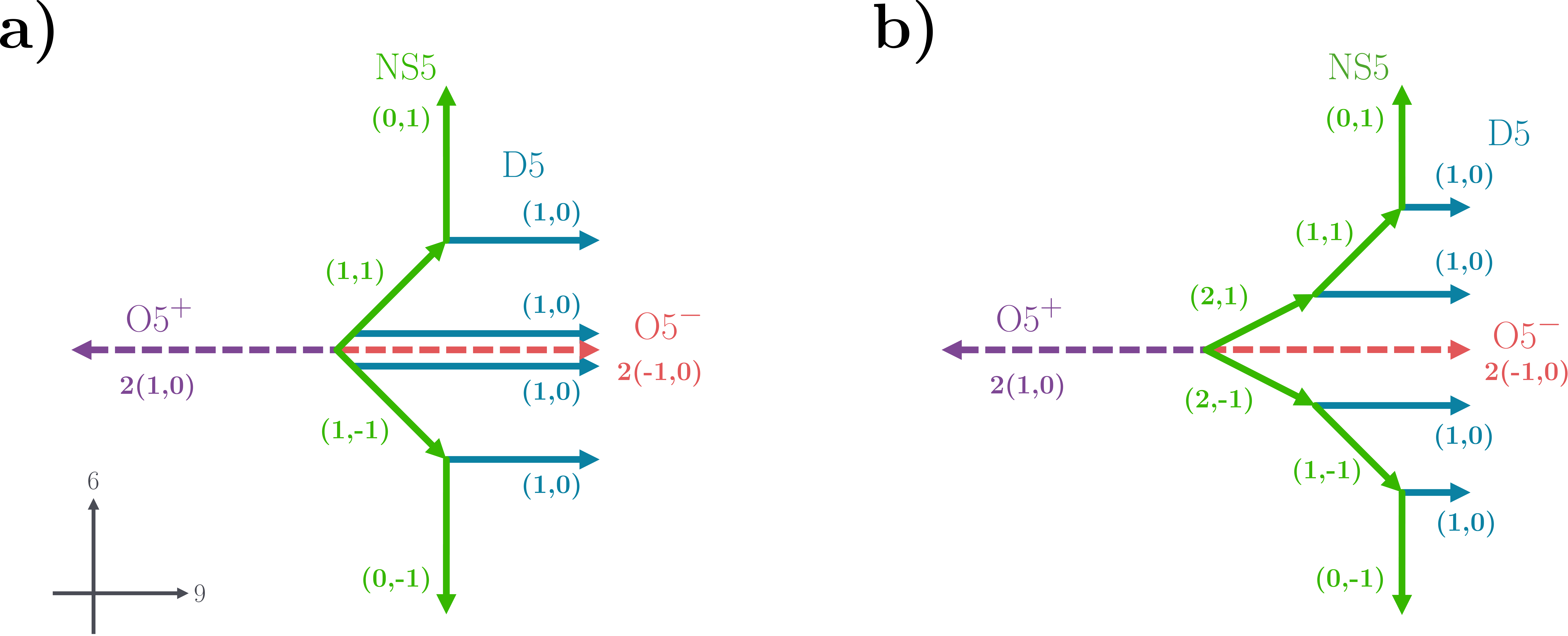}
\caption{\small The fork  configuration for one (a) or two (b) generic locations for the D5'-branes (and their images). The configuration turns into a non-trivial $(p,q)$ web due to the intermediate $(p,q)$ segments.}
\label{fig:masses-zero}
\end{center}
\end{figure}

We will refer to these configurations in which the O5'-planes change sign and there are extra D5'-brane charges as `forks', because of their graphical representation (c.f. \cite{Brunner:1998jr}). We can deform the above simplest fork configuration by moving the half-D5'-brane in pairs away from the O5'-plane. This will imply the the NS5-brane acquires some D5'-brane charge and becomes a more general $(q,1)$ 5-brane, see Figure \ref{fig:masses-zero}. Specifically, by moving one D5'-brane and its image away from the O5'$^-$-plane, the $(0,1)$ 5-brane becomes a $(1,1)$ 5-brane, and the $(0,-1)$ 5-brane becomes its orientifold image $(1,-1)$ 5-brane. Similarly, moving two D5'-branes and their two images away from the O5'$^-$-plane (hence leaving it empty), the NS5-brane halves turn into $(2,1)$ and $(2,-1)$ 5-branes, respectively.

\begin{figure}[htb]
\begin{center}
\includegraphics[scale=.075]{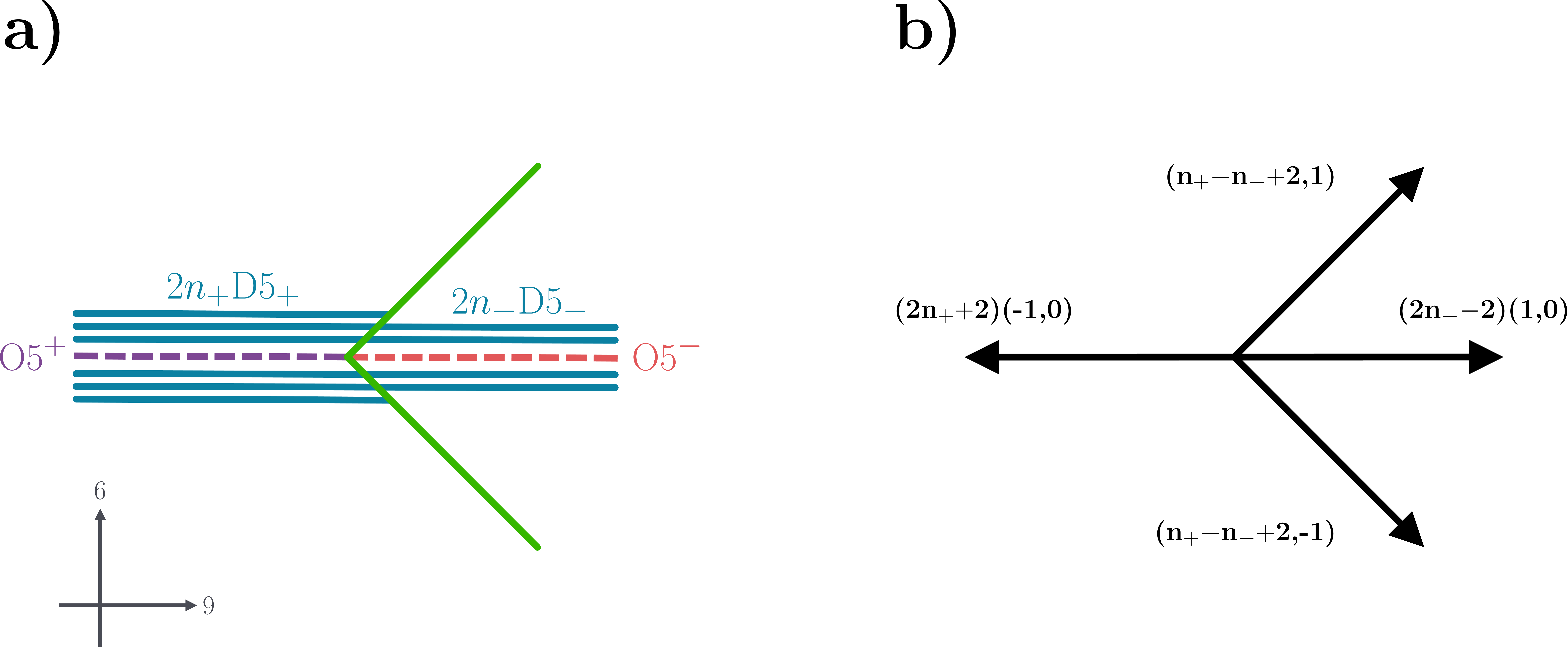}
\caption{\small A general fork  configuration for $C_0=0$. In (a) we draw the objects and in (b) the flow of 5-brane charges. }
\label{fig:fork-general-0}
\end{center}
\end{figure}

We can consider even more general configurations, which are obtained from the latter by adding a general number $2n_+$ half D5'-branes on top of the O5'$^+$-plane (denoted as D5'$_+$-branes) and $2n_-$ half- D5'-branes on top of the O5'$^-$-plane (D5'$_-$-branes). The fact that the number of half D5'$_+$-branes must be even is because the orientifold projection acts with an antisymmetric matrix on the D5'$_+$-brane Chan-Paton indices; equivalently, the gauge group on the D5'$_+$-brane worldvolume is $USp(2n_+)$, so we need an even number. The fact that the number of half D5'$_-$-branes is also even follows from charge conservation and the way the orientifold acts on charges for $C_0=0$, as follows. On the O5'$^+$-plane side, the total RR charge is $2n_+ + 2$, whereas on the O5'$^-$-plane side, the RR charge is $2n_--2$. The difference is $2n_+-2n_-+4$ and this has to be absorbed by the NS5-brane, namely the $(0,1)$ 5-brane turns into a $(n_+-n_-+2,1)$ 5-brane and the $(0,-1)$ 5-brane turns into a $(n_+-n_-+2,-1)$ 5-brane. The $(n_+-n_-+2,1)$ 5-brane and the $(n_+-n_-+2,-1)$ 5-brane are related by the O5'-plane in a compatible way for $C_0=0$. The symmetry of the configuration thus requires the number of D5'$_-$-branes to be even as well. The resulting configuration is shown in Figure \ref{fig:fork-general-0}.

Let us pose for a moment in this configuration and discuss the symmetries and the 5d matter spectrum at the intersection. The D5'$_+$-branes have a worldvolume symmetry $USp(2n_+)$ and the D5'$_-$-branes have a symmetry $SO(2n_-)$. These symmetries propagate on a half 6d space, but they have a localized coupling to the 5d matter at the intersection\footnote{If necessary, it is possible to turn the whole gauge sector into a 5d one, by making the extent in $x^9$ compact by adding extra 5-branes to end the D5'-branes, as in 5d gauge theory Hanany-Witten brane configurations \cite{Aharony:1997ju,Aharony:1997bh}. We will not do so explicitly but advise the reader to keep this possibility in mind in case of need of extra 5d intuition.\label{foot-interval}}. In particular, at the intersection of the D5'-branes there is one half-hypermultiplet charged under both gauge factors in the bifundamental $(2n_+,2n_-)$. From the perspective of the $USp(2n_+)$ the fact that there is an even number $2n_-$ of flavours in the fundamental is of great importance, because it implies that there is no localized contribution to the 5d global gauge anomaly \cite{Intriligator:1997pq} associated to %
\beqa
\Pi_5(USp(2n))=\IZ_2\, .
\label{pi5}
\eeqa
This potential (but absent in this case) anomaly will come back in an even more interesting way in the $C_0=1/2$ case in section \ref{sec:co12}.

The introduction of D3-branes ending on the above 5-brane configurations is discussed in section \ref{sec:fork-bc-parity-anomaly}, while the gravity duals of the 4d $\NN=4$ SYM theory on half-space with boundary conditions defined by the above 5-brane configurations are described in section \ref{sec:d3-ns5-d5-o5-dual}.

\subsubsection{O5'-plane configurations at $C_0=1/2$, and 5d global gauge anomaly}
\label{sec:co12}

Let us now repeat a similar discussion, but for the other possible orientifold invariant value of the IIB axion, namely $C_0=1/2$. Namely, we are discussing the tilded variants of O5-planes, i.e. ${\widetilde{\rm O5'}}{}^\pm$-planes, introduced in section \ref{sec:o5}, but for simplicity, we abuse notation and continue to refer to them as O5'$^\pm$-planes, hoping no confusion arises.

\begin{figure}[htb]
\begin{center}
\includegraphics[scale=.075]{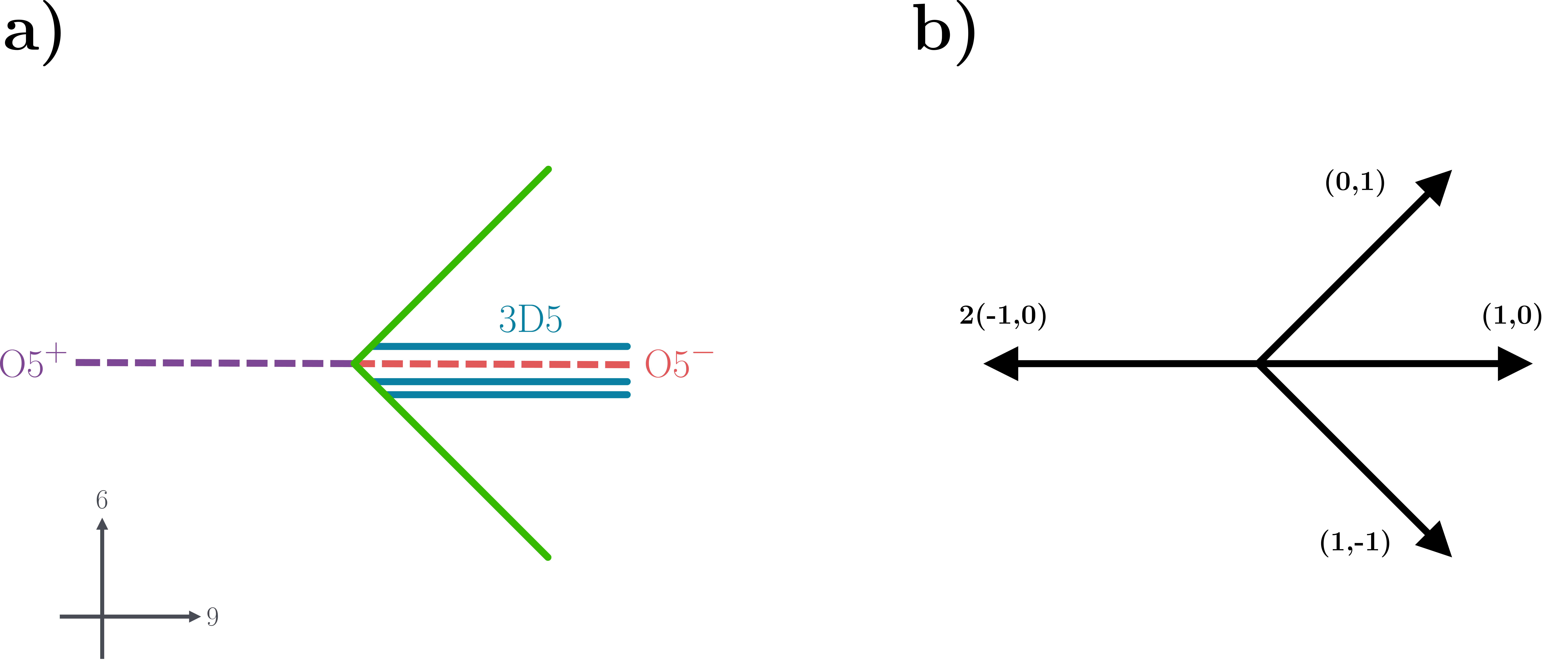}
\caption{\small A simple fork  configuration for $C_0=1/2$. In (a) we draw the objects and in (b) the flow of 5-brane charges. }
\label{fig:simplest-fork-12}
\end{center}
\end{figure}

\begin{figure}[htb]
\begin{center}
\includegraphics[scale=.095]{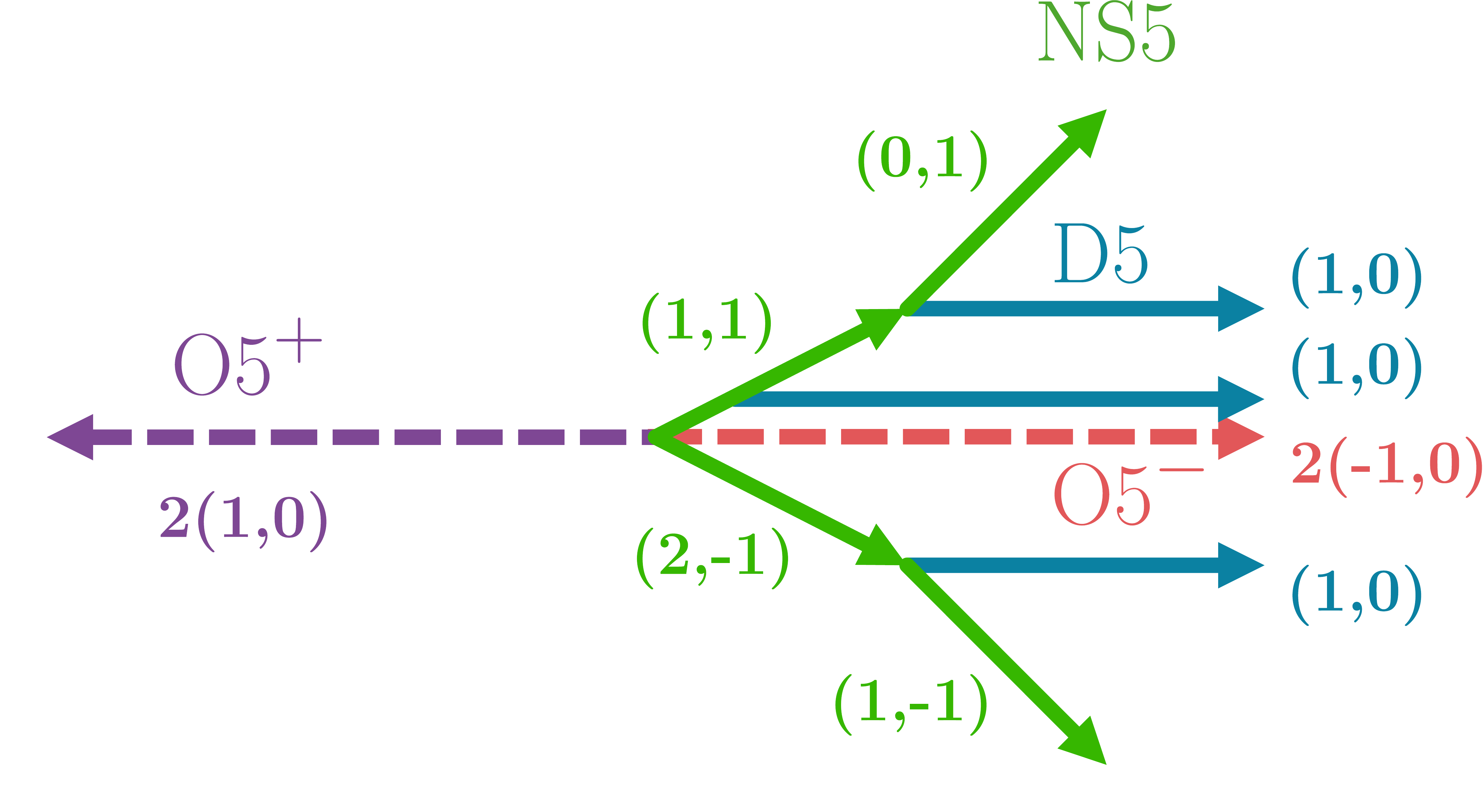}
\caption{\small The fork  configuration with $C_0=1/2$ for a generic location for a D5'-branes (and its image).}
\label{fig:masses-12}
\end{center}
\end{figure}

\begin{figure}[htb]
\begin{center}
\includegraphics[scale=.075]{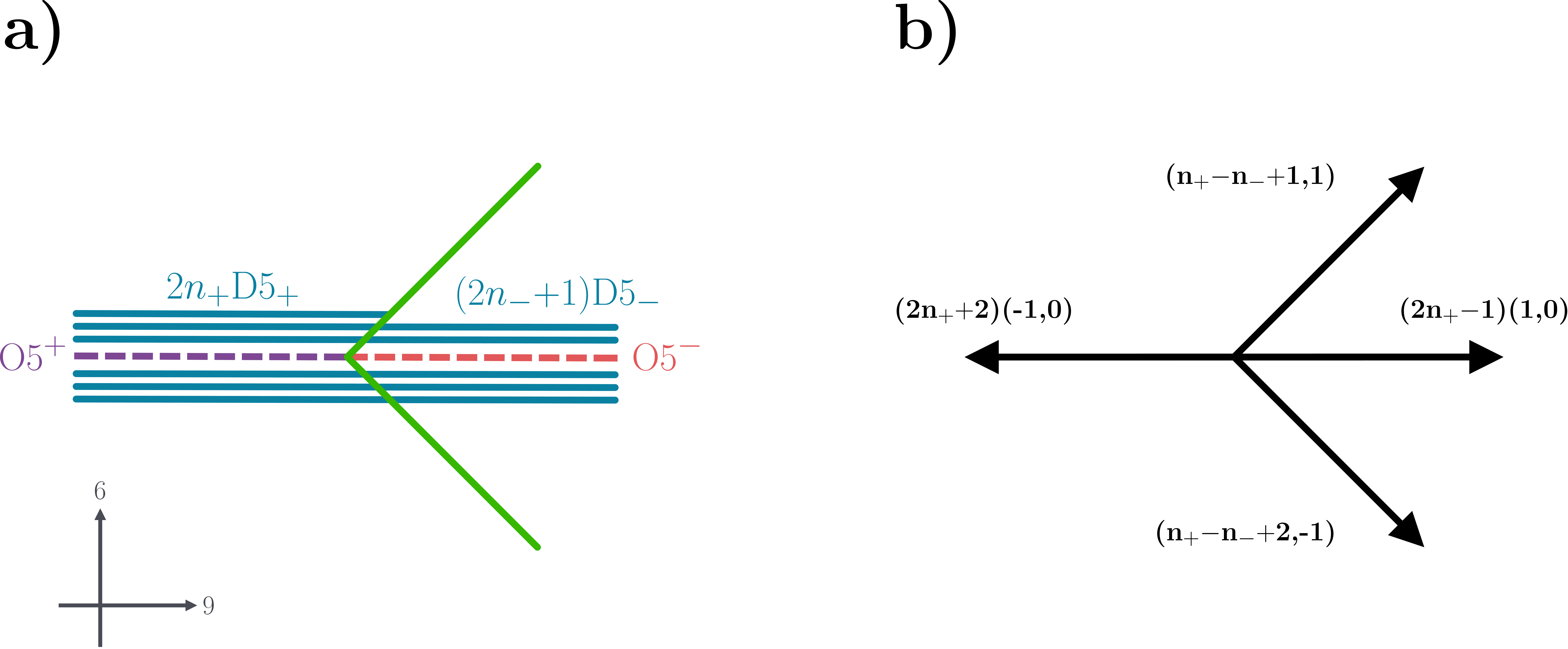}
\caption{\small A general fork  configuration for $C_0=1/2$. In (a) we draw the objects and in (b) the flow of 5-brane charges. }
\label{fig:fork-general-12}
\end{center}
\end{figure}

As mentioned above, now a $(p,q)$ 5-brane goes in the direction $z=p+\tau q$ in the 96 plane, namely $(x^9,x^6)=(p+q/2,q/g_s)$. Since the orientifold flips the coordinate $6$, such a $(p,q)$ 5-brane will be mapped to a 5-brane going in the direction $(x^9,x^6)=(p+q/2,-q/g_s)$, namely it corresponds to a $(p+q,-q)$ 5-brane. For instance, charge vector $(1,0)$ is invariant, but the charge vector $(0,1)$ is exchanged with the charge vector $(1,-1)$. In other words, the action of the orientifold on the $(p,q)$ charges is via the matrix
\beqa
\begin{pmatrix}
1 & 1 \cr 0 & -1
\end{pmatrix}\, .
\eeqa
This implies that the structure of the fork is different, and in particular it requires that we have an odd number of half D5'-branes on top of the O5'$^-$-plane (given that we still need an even number of D5'-branes on top of the O5'$^+$-plane). This fits perfectly with the fact that, as mentioned, the O5'$^\pm$-planes for $C_0=1/2$ actually correspond to ${\widetilde{\rm O5'}}^\pm$-planes, and the ${\widetilde{\rm O5'}}^-$-plane is just an O5'$^-$-plane with a stuck D5'-brane. A simple example of such fork is shown in Figure \ref{fig:simplest-fork-12}, and the forks for generic positions for 5-branes are shown in Figure \ref{fig:masses-12}. The general configuration with D5'-branes only on top of the O5'-planes, i.e. with $2n_+$ half D5'$_+$-branes and $2n_-+1$ D5'$_-$-branes is shown in Figure \ref{fig:fork-general-12}.

Let us now consider the symmetries and matter content on the 5-branes of the configuration. We have $USp(2n_+)\times SO(2n_-+1)$ gauge symmetries propagating on the corresponding 6d half-spaces, and a localized 5d bifundamental half-hypermultiplet $(2n_+,2n_-+1)$. In contrast with the $C_0=0$ case, this implies that for the $USp(2n_+)$ there is an odd number of flavours in the fundamental, and hence a localized contribution to the 5d $\IZ_2$ global gauge anomaly characterized by (\ref{pi5}).

\medskip

{\bf Global gauge anomaly inflow}

It is important to recall that the gauge symmetries live on a 6d half-space, and that the matter is localized on a 5d slice. This means that there could be extra contributions from the anomaly from other flavours located elsewhere in the direction 9 and the $USp$ theory could still be consistent, recall footnote \ref{foot-interval}. This is for instance the case if we locate another NS5-brane (possibly with some induced D5'-brane charge) and its orientifold image, at some position $x^9<0$, so that the D5'$_+$ branes now live on a segment and give a genuine 5d $USp(2n_+)$ theory. Then, because of the introduction of the new NS5-brane, the O5'$^+$-plane would flip to another O5'$^-$-plane across it, and the new piece of O5'$^-$-plane would come with an odd number of D5'-branes on it. This would introduce another odd number of flavours for the $USp(2n_+)$, making the total number of flavours even and cancelling the $\IZ_2$ global gauge anomaly. However, it is important that the two sets of flavours are located at different positions in $x^9$, so there must be some inflow of anomaly at these points, which cancels globally when we consider the whole segment.

Let us then come back to the case of the semi-infinite D5'$_+$-branes and discuss the nature of the inflow\footnote{We are grateful to I\~naki Garcia-Etxebarria and Miguel Montero for very useful discussions on  global gauge anomaly inflows.} of global anomaly from the 6d bulk of the D5'$_+$-branes into the 5d boundary defined by the 5-brane intersections. In order to describe an inflow, one simply needs to describe the 6d anomaly theory, namely a 6d topological action on a 6d space whose boundary is the 5d manifold where the anomaly is localized. This kind of inflow has been considered from the mathematical perspective in \cite{Witten:2019bou}, and can be understood from the perspective of the Dai-Freed theorem \cite{Dai:1994kq} (see \cite{Garcia-Etxebarria:2018ajm} for a recent physicists' account), which gives the 6d anomaly theory in terms of the corresponding $\eta$ invariant 
\beqa
S_{DF,6}=\frac 12 \int_{6d_+} \eta_6\, ,
\label{dai-freed}
\eeqa
whose contribution on manifolds with boundary is precisely the partition function of an odd number of fermions in the fundamental representation, thereby cancelling the contribution from the explicit fermion spectrum. Hence, this action should actually arise on the world-volume of the D5'$_+$-branes on top of the O5'$^+$-plane. We are not aware of an explicit derivation of this coupling, hence in the following we will assume it to be present, as requires to have a consistent configuration. From a heuristic derivation, this coupling should be regarded as a K-theory generalization of the familiar Chern-Simons coupling 
\beqa
\int_{6d_+} C_0 \tr F^3
\eeqa
of the parent theory on the D5'$_+$-branes before the orientifold quotient, which freezes the value $C_0=1/2$ and morally turns the Chern class into a $\IZ_2$ valued index for the $USp(2n_+)$ theory. Note that this heuristic explanation makes it manifest that the anomaly inflow is absent in the $C_0=0$ case of section \ref{sec:co0}, nicely matching the fact that the number of fermion flavours is even in that case, so there is no global gauge anomaly to cancel.

The above is to our knowledge the first physical realization of the global gauge anomaly cancellation inflow of \cite{Witten:2019bou} in string theory. It would be interesting to explore this kind of effect in other setups. As example, we provide in appendix \ref{sec:4d-global-anomaly} a similar system involving inflow for Witten's global gauge anomaly for 4d $USp(n)$ with an odd number of fermions in the fundamental.

The fact that the $C_0=1/2$ configurations lead to an interesting pattern of symmetries and anomalies will continue upon the introduction of semi-infinite D3-branes ending in the fork, to which we turn next.

\subsection{D3-branes ending on O5'-plane forks. The 3d parity anomaly}
\label{sec:fork-bc-parity-anomaly}
 
In this section we consider the NS5/D5'/O5'$^\pm$ fork configurations as providing the boundary conditions for a stack of D3-branes. We consider $N$  D3-branes extending in 012 and of semi-infinite extent in $x^3>0$, ending at $x^3=0$ at the intersection of the NS5/D5'/O5' system; actually, because we have an O5'-plane which flips the direction 3, there is the orientifold image stack of $N$ D3-branes at $x^3<0$. Overall, we have a stack of $N$ D3-branes along 0123, which is split in two semi-infinite halves by the NS5/D5/O5 system at $x^3=0$.

The 5-brane sector of the story is as discussed in previous sections, so we focus on the new information associated to the D3-branes. Part of the discussion is analogous to section \ref{sec:o5}. In particular, the arguments motivating that we just need to focus on the region close to the orientifold planes hold similarly. We thus focus on a system of $n$ D3-branes, intersecting with the 5-brane fork configuration containing the O5'-planes. The gauge symmetry on the D3-branes is $SU(n)$, because the D3's at $x^3>0$ are mapped not to themselves but to the D3's at $x^3<0$, so we have a 4d $\NN=4$ $SU(n)$ SYM defined on a half-space. The spectrum of degrees of freedom localized at $x^3=0$ can be computed using standard techniques for analogous brane configurations, see e.g. \cite{Hanany:1997sa,Evans:1997hk,Landsteiner:1998gh,Brunner:1998jr,Elitzur:1998ju,Park:1999eb,Park:1999ep}, and \cite{Giveon:1998sr} for a review. We discuss the $C_0=0,1/2$ cases separately.

\medskip

{\bf The $C_0=0$ case}

Let us start with the configurations with $C_0=0$ for simplicity, and consider the configuration in which the O5$^+$-plane turns into an O5$^-$ plus 4 D5'-branes, and they are separated by just an NS5-brane. Hence the 5-brane configuration does not involve non-trivial $(p,q)$ 5-branes. There is a sector of 3d localized modes arising from the open strings stretched among the D3-branes and their orientifold images. As discussed in \cite{Landsteiner:1998gh,Brunner:1998jr,Elitzur:1998ju} in related setups for 4d gauge theories, the 3d $\NN=4$ bifundamental hypermultiplet splits into two 3d $\NN=2$ chiral multiplets, with each one of them feeling a different orientifold projection. This leads to $\NN=2$ chiral multiplets in the $\symm+\antiasymm$. In addition, there is a sector of 3d localized modes from the open strings between the D3-branes and the half D5'-branes, leading to 4 3d $\NN=2$ chiral multiplets in the $\antifund$.

It is now easy to understand the spectrum in the slightly more involved configurations for $C_0=0$ when the half D5'-branes are moved away from the O5'$^-$-plane and the configuration turns into a non-trivial $(p,q)$ web, recall Figure \ref{fig:masses-zero}. This simply corresponds to turning on 3d real masses (i.e. masses corresponding to background values for 3d vector multiplets \cite{Aharony:1997bx}) for the flavours in the D3-D5' sector. If we move one D5'-brane and its orientifold image away from the O5'$^-$-plane, we are left with just 2 chiral multiples in the $\antifund$, and the D3-branes are ending on the intersection of the O5'$^\pm$-plane with one $(1,1)$ and one $(1,-1)$ 5-branes. If we move two D5'-branes and their orientifold images, we are left with no flavours, and the D3-branes are ending on the intersection of the O5'$^\pm$-plane with one $(2,1)$ and one $(2,-1)$ 5-branes.

An important observation at this point is that the field theory analysis implies that when 3d $\NN=2$ flavours become massive due to real masses and are integrated out, they generate a 1-loop Chern-Simons term. For instance, for the case of purely abelian symmetries $U(1)_i$ with gauge bosons $A_i$, the terms arising from integrating out fermions (labeled by $f$), of charges $(q_f)_i$ and masses $m_f$, have the structure
\beqa
k_{ij}\int_{3d} A_i dA_j\quad ,\quad k_{ij}=\frac 12\sum_f (q_f)_i(q_f)_j {\rm sign}(m_f)\, .
\eeqa
For non-abelian symmetries, there is a similar expression involving the Chern-Simons 3-form $\omega_3(A_i)$ defined by $d\omega_3=\tr F^2$, and with the combination of the charges replace by $d_{3}(R_f)$, the cubic Casimir of the representation of the fermion $f$. We also note that the above structure of Chern-Simons arises also for global symmetries, if the fermions carry charges under them, by simply interpreting the corresponding $A_i$ as a background field. 

In the brane realization the appearance of Chern-Simons terms is encoded in the fact that D3-branes are ending not on NS5-branes, but rather on $(p,1)$-branes. It was shown in \cite{Bergman:1999na} that the boundary conditions that such 5-branes impose on the gauge fields on the D3-branes precisely reproduce this Chern-Simons term. In the $C_0=0$ case, the $(p,q)$ charges are such that the coefficients of the Chern-Simons terms are integer, in agreement with the standard quantization condition for Chern-Simons levels.  In other words, we can start in the configuration of four massless chiral multiplets, in which the 5-brane charge is continuous across the fork, so the two opposite charge O5'-planes are separated by an actual NS5-brane, in which case there is no Chern-Simons term. We can then make flavours massive by moving D5'-branes off the O5'$^-$-plane in pairs, and this doubling implies that the 1-loop Chern-Simons term has integer coefficient, hence matching the coefficient of the Chern-Simons term as directly read from the now non-trivial $(p,q)$ charges of the 5-branes at the endpoint of the D3-brane. 

This $C_0=0$ configuration can be enriched by adding additional equal numbers of $n_\pm$ half D5'$_\pm$-branes, either on top of the O5'$^\pm$-planes, or away from them. The result corresponds to introducing additional chiral multiplets in the fundamental or antifundamental representation, possibly with real mass parameters describing their location in $x^6$. The resulting $(p,q)$ web reproduces the necessary Chern-Simons terms on the D3-branes required if such masses are turned on. 

\medskip

{\bf The $C_0=1/2$ case: The 3d parity anomaly}

Consider now the $C_0=1/2$ configurations, which are similar to the previous ones with one important difference. Indeed we obtain 3d localized modes corresponding to 3d $\NN=2$ chiral multiplets in the $\symm+\antiasymm$ from the D3-D3 open string sector, plus a number of $\antifund$'s from the D3-D5'$_-$ sector. The key difference lies in the fact that in this case there is necessarily an odd number of half D5'$_-$-branes on top of the O5'$^-$-plane, which results in an odd number of 3d $\NN=2$ chiral multiplets in the (anti)fundamental representation. This leads to the celebrated 3d parity anomaly \cite{Redlich:1983kn,Redlich:1983dv,Alvarez-Gaume:1984zst}, which is realized in our setup as follows. The odd number of chiral multiplets would lead to a lack of invariance under large gauge transformations, but which in 3d can be removed by the introduction of a 3d Chern-Simons term of the kind discussed above, but with half-integer coefficient. This is precisely the Chern-Simons term which arise on the D3-branes because of the  $(p,q)$ 5-brane configuration on which they are ending, which is necessarily non-trivial for $C_0=1/2$. For instance, considering the case of one O5'$^+$-plane turning into one O5'$^-$-plane and 3 half D5'$_-$-branes, the D3-branes end on the intersection of the O5'-planes with a $(0,1)$ 5-brane and a $(1,-1)$ 5-brane. This gives the half-integer coefficient for the Chern-Simons, precisely necessary to cancel the 3d parity anomaly of the odd number of flavours. The presence of a 3d Chern-Simons term can be read in field theory as a boundary term of the 4d $\NN=4$ $SU(N)$ theory because of the non-trivial $\theta=\pi$ term at $C_0=1/2$.

Other examples can be worked out similarly, e.g. separating one D5'-brane and its image (recall Figure \ref{fig:masses-12}), or general numbers of D5'$_\pm$-branes (recall Figure \ref{fig:fork-general-12}), with the result that there is always a Chern-Simons term with half-integer level, as required to ensure gauge invariance. On the other hand, the Chern-Simons terms breaks 3d parity, displaying a new interesting symmetry breaking pattern due to the reduced supersymmetry of the boundary conditions in our system.

For later convenience, let us express the structure of parity-violating Chern-Simons couplings in a way easier to recognize from the viewpoint of the gravitational dual. Recall from the field theory argument above that the global gauge anomaly arising from the odd number of fermions also affects the global symmetries under which such fermions are charged, in particular the $SO(2n_-+1)$ global symmetry. This implies that the structure of the Chern-Simons coupling, in a simplified abelian case, is morally of the form
\beqa
\frac 12 \int_{3d} (A_{\rm D3}-A_{\rm D5'})\, d(A_{\rm D3}-A_{\rm D5'})\, ,
\label{crude-cs}
\eeqa
where $A_{\rm D3}$ is the D3-brane worldvolume gauge field, and $A_{\rm D5'}$ is the background field for the corresponding global symmetry from the D3-brane worldvolume perspective, in other words, the 6d D5'$_-$-brane worldvolume gauge field. 
Note that the coupling effectively involves the difference of gauge fields on D3- and D5'-branes. This is because D3-branes ending on D5-branes have Dirichlet boundary conditions for vector multiplets. Hence the gauge bosons on both kinds of D-branes are effectively locked, in other words they couple collectively in a similar way. 

Hence, the presence of the above couplings can be identified from the perspective of the D5-branes as the presence of a localized coupling for the 6d gauge fields
\beqa
\frac n2 \int_{3d}\omega_3(A_{\rm D5'})\, ,
\label{parity-D5}
\eeqa
where $\omega_3$ is the Chern-Simons 3-form, valid also in the non-abelian case, and we have added a factor of $n$ to account for the fermion multiplicity. This provides a simple way to test the existence of the couplings (\ref{crude-cs}) using the D5'-branes without explicit reference to the D3-branes.

Hence, we have found a very interesting structure of symmetries and anomalies for configurations of D3-branes ending on O5'-planes dressed up with 5-branes, arising from the 3d $\NN=2$ supersymmetry of the local configuration.

As a final remark, let us again emphasize that we have focused on the classification of configurations near the O5'-planes, and that each such configuration can be combined with additional 5-branes away from it (and their orientifold images) to produce large classes of $1/4$ supersymmetric orientifold boundary conditions for 4d $\NN=4$ $SU(N)$ SYM. In the next section we turn to a discussion of their gravity duals.

\section{Gravity duals: ETW branes, symmetries and anomalies}
\label{sec:d3-ns5-d5-o5-dual}

We now take the limit of large number of D3-branes and replace them by their gravity dual. The approach is in analogy with section \ref{sec:o5-dual}, and we focus on the gravity dual of the configuration near the orientifold plane, ignoring the possible presence of extra 5-branes away from it. We also ignore the backreaction of the O5'-planes, and of the possible NS5- or D5'-branes on top of it. Hence, we must simply characterize orientifolds of  AdS$_5\times \IS^5$, and add the configuration of 5-branes as probes\footnote{We however note that, if necessary, it is possible to allow for backreaction of the NS5-brane charge, using the solutions in section \ref{sec:d3-ns5-d5-dual}, with the appropriate orientifolding and introduction of D5'-branes if present; this is a good approximation in the weak coupling regime $g_s\ll 1$, where the backreaction of D5'-brane charges is negligible compared with that of NS5-branes. In the graphical representation of the $(p,q)$ webs, this limit corresponds to elongating along the vertical direction, so that any $(p,q)$ 5-brane essentially aligns with the vertical direction.}.

\subsection{The O5'-plane orientifold of AdS$_5\times\IS^5$}
\label{sec:gravity-dual-o5prime}

In this section we describe explicitly the orientifold quotient of AdS$_5\times\IS^5$ appropriate to describe O5'-planes. Recall, that the O5'-plane spans the directions 012 459, so the orientifold quotient implies the action $\Omega R'$, with $R'$ flipping the directions 3678, namely $(x^3,x^6,x^7,x^8)\to(-x^3,-x^6,-x^7,-x^8)$ in the parent flat space. The action on the sphere can be read out from the embedding of the unit ball in $\IR^6$
\beqa
(x^4)^2+(x^5)^2+(x^6)^2+(x^7)^2+(x^8)^2+(x^9)^2=1\, .
\eeqa
The flip in the directions 678 means that the O5'-plane is wrapped on the $\IS^2$
\beqa
\IS^2_{\rm {O5'}}\,:\; (x^4)^2+(x^5)^2+(x^9)^2=1\, .
\label{s2o5}
\eeqa

Since the orientifold flips the direction 3, the O5'-plane (and the possible 5-branes on top of it) play the role of ETW boundary in this configuration. 

Let us emphasize that this $\IS^2$ is not any of the $\IS^2$'s wrapped by the NS5- and D5-branes in the configurations preserving 8 supersymmetries. Indeed, let us discuss the interplay of the new sphere with the old ones. The NS5-brane spans the directions 012456 and is located at the origin in 3789 in flat space. This means that it fills out the AdS$_4$ and wraps the $\IS^2$ given by
\beqa
\IS^2_{\rm NS5}\,:\; (x^4)^2+(x^5)^2+(x^6)^2=1\, .
\label{s2ns5}
\eeqa
The $\IS^2$ spanned by the NS5 and the $\IS^2$ spanned by the O5'-plane intersect over the $\IS^1$
\beqa
{\rm Inters:}\, (x^4)^2+(x^5)^2=1\, .
\label{s1}
\eeqa
This mean that the NS5 splits the O5'-plane in two halves, each wrapped on each of the two hemispheres of the $\IS^2_{\rm O5'}$ in (\ref{s2o5}), corresponding to (\ref{s2o5}) for $x^9\lessgtr$ 0. We denote these two hemispheres (topologically, disks) by $\ID^2_\pm$, and note that, as in the flat space discussion, they are wrapped by opposite RR charge O5'$\pm$-planes. Conversely, the O5-plane splits the NS5-brane in two halves, so that the corresponding two hemispheres of the $\IS^2$ in (\ref{s2ns5}) correspond to different $(p,1)$ 5-branes, as befits the fork configuration. Note that, because the $(p,1)$ 5-brane is bent in the 96 plane, the actual hemisphere is not exactly\footnote{In case a more precise description is required, we note that the $(p,q)$ 5-brane goes in the direction $z=p+\tau q$ namely $x^9=(p+C_0 q)t$, $x^6=(q/g_s) t$, with $t\in[0,\infty)$, so the hemisphere is squashed to
\beqa
(p,q):\, (x^4)^2+(x^5)^2+[(p+C_0 q)^2+(q/g_s)^2]\,t^2=1\quad t>0\, .
\label{s2ns5f}
\eeqa
Again, the topological structure is as in the simplified description, which is in fact recovered for $g_s\to 0$.} given by (\ref{s2ns5}), but topologically we have the same structure. 

Let us finally note that we can add D5'$_\pm$-branes on top of the O5'$^\pm$, namely they span AdS$_4\times \ID^2_\pm$, and use arguments about continuity of RR charges to recover the same fork configurations studied in the flat space context. The spectrum on the worldvolume of the D5'-branes can hence be analyzed as in section \ref{sec:fork-c0}. We choose to discuss together the cases $C_0=0,1/2$, by introducing $2n_+$ D5$_+$-branes and $2n_-$ or $2n_-+1$ D5$_-$-branes, for $C_0=0$ or  $C_0=1/2$. We obtain $USp\times SO$ gauge groups on the 6d spaces AdS$_4\times \ID^2_\pm$, which have the  $\IS^1$ in (\ref{s1}) as boundary, and which supports a 5d half-hypermultiplet in the bifundamental $(\fund,\fund)$. For the case $C_0=0$ there is no local contribution to the $\IZ_2$ 5d global gauge anomaly, but for $C_0=1/2$ the number of 5d fermions in the fundamental of $USp(2n_+)$ is even and there is a non-trivial local contribution to the $\IZ_2$ 5d global gauge anomaly. This should be cancelled by the same global gauge anomaly inflow mechanism discussed in section \ref{sec:co12}, with the 6d space supporting the coupling (\ref{dai-freed}) is AdS$_4\times \ID^2_+$.

As explained, these geometries describe ETW configurations for an asymptotic AdS$_5(\times\IS^5)$. An important step in this interpretation is to perform the dimensional reduction to a 5d bulk with an 4d ETW boundary. We may expect that the reduced amount of supersymmetry in the system can lead to a rich structure of worldvolume fields on the ETW boundary.
In the coming section we will see that there are important subtleties in these statements.

\subsection{SymTFT, a puzzle and its resolution}
\label{sec:puzzle}

In this section we explore the SymTFT interpretation of the above background by performing the topological reduction of the systems to a 5d bulk with 4d boundary conditions. For $C_0=1/2$, this seemingly leads to a puzzle due to a clash between the rich higher-dimensional phenomena occurring on the ETW configuration and the far simpler topological structure requires by SymTFT couplings in the lower-dimensional picture. The resolution of the puzzle will motivate  more general lessons which we explore from a broader perspective in section \ref{sec:lesson}.

\subsubsection{A puzzle in the D5'-brane open string spectrum}
\label{sec:puzzle-d5}

Let us consider the reduction of the above configurations to AdS$_5$ with an AdS$_4$ ETW boundary. This is basically the setup arising in the derivation of the SymTFT of the 4d $\NN=4$ SYM theory on half-space, with boundary conditions defined by the O5'-plane configurations. Using arguments familiar by now, we mostly focus on the reduction of the objects located on top of the AdS$_4$ ETW brane, i.e. the location of the O5'-plane.

Recall that on the $\IS^5$ of the double cover at this location, we have $2n_+$ D5'$_+$-branes and $2n_-+1$ D5'$_-$-branes wrapped on the hemispheres $\ID^2_\pm$, leading to a 6d $USp(2n_+)\times SO(2n_-+1)$ gauge sector, with a 5d bifundamental half-hypermultiplet on the equator $\IS^1$ at which they meet with the hemispheres wrapped by the NS5-branes (or $(q,1)$ 5-branes, more in general). Upon reduction on the internal space, we get a 4d $USp(2n_+)\times SO(2n_-+1)$ gauge symmetry in AdS$_4$, seemingly still with a 4d bifundamental half-hypermultiplet, arising from the redution of the 5d half-hypermultiplet. This spectrum however leads to a 4d $USp(2n_+)$ with an odd number of fermions in the fundamental representation, leading to a Witten 4d global gauge $\IZ_2$ anomaly.

One may be tempted to expect this anomaly to be solved via a global anomaly inflow, just like in the original higher dimensional setup. However, this is not possible in the present case. The only possible source of inflow would be the AdS$_5$ bulk, but the 5d bulk topological couplings are those of the SymTFT of the 4d $\NN=4$ SYM theory, c.f. (\ref{bf-symtft}), which do not lead to an anomaly inflow of that kind.

One might entertain the possibility that perhaps such 5d anomaly inflow topological couplings exist and have been overlooked in the literature. But it is easy to argue that this cannot be the case. The basic argument is that, if such topological couplings existed, they would lead to a global anomaly inflow in any boundary of the bulk theory. The argument is similar to the universal appearance of edge modes in topological superconductors because bulk topological couplings necessarily lead to an inflow towards any boundary of the chunk of material.

In the present context it is easy to discuss other 4d boundaries of the bulk 5d theory, and check no such inflow is present, as can be argued from different perspectives. 

$\bullet$ {\bf Holographic boundary:} Most prominently, the 4d physical boundary of the SymTFT, which is given by the 4d $\NN=4$ SYM on the 4d holographic boundary, does not suffer from any 4d $\IZ_2$ global gauge anomaly, so no 5d bulk couplings of that kind should exist. Note that this argument is topological, so it holds even though this is a holographic boundary, hence located at infinite distance. In any event, the next examples involve boundaries at finite spacetime distance.

$\bullet$ {\bf Additional ETW branes in wedge holography:} It is possible to consider the holographic setup with a second AdS$_4$ ETW brane in AdS$_5$, a setup known as wedge holography \cite{Akal:2020wfl,VanRaamsdonk:2021duo}, which provides the gravity dual of a 4d CFT on an interval with boundary conditions given by two independent BCFT$_3$'s. In our setup we have D3-branes stretched between the $C_0=1/2$ fork and an arbitrary set of 5-branes away from it, which may not involve fork configurations, see section \ref{sec:lesson} for explicit examples. Hence, the fact that there exists a second boundary which is consistent without an inflow from the 5d bulk implies that such 5d couplings are absent and cannot assit the ETW fork boundary with any inflow.

$\bullet$ {\bf Bag holography:} A similar argument can be run by starting with the basic fork configuration, we can add additional $(p,q)$ 5-branes near but slightly away from it. As discussed in \cite{Huertas:2023syg}, there exists a scaling limit, which essentially amounts to shutting off the semi-infinite D3-branes, in which the asymptotic AdS$_5\times\IS^5$ is shut off and one is left with AdS$_4\times X_6$ now with  $X_6$ a fully 6d compact space. In other words, one effectively includes a second boundary in the configuration, associated to 5-branes away from the fork. Since the latter do not lead to global gauge anomalies, we can run the above argument and conclude that there cannot be 5d bulk couplings leading to a global gauge anomaly inflow for any boundary.

\subsubsection{Resolution of the puzzle}
\label{sec:puzzle-resolved}

The resolution of the above puzzle is actually simple. The key point is that the reduction of the 5d bifundamental half-hypermultiplet requires some care, and that assumption that it produces a 4d bifundamental half-hypermultiplet is actually incorrect. Indeed, this would be the case if the compactification would be with periodic boundary conditions for the fermions. But in our configuration the $\IS^1$ is contractible, or more precisely it is the boundary of a 2-chains given by the hemispheres $\ID^2_\pm$ (also, by the hemispheres of wrapped by the $(p,1)$ 5-branes), and this determines that fermions are actually antiperiodic, as we now argue. 

The argument is actually related to the anomaly inflow of the 6d $USp(2n_+)$ bulk into its 5d boundary. Recall we start out with a bifundamental half-hypermultiplet on AdS$_4\times\IS^1$, which leads to a global gauge anomaly due the odd number of 5d fermions in the fundamental of the $USp(2n_+)$ theory. This must be cancelled by the inflow from the 6d bulk of the $USp(2n_+)$ D5$_+$-brane worldvolume on AdS$_4\times\ID^2_+$ (with $\partial \ID^2_+=\IS^1$). But for this mechanism to work, one must have specific boundary conditions (APS boundary conditions) \cite{Dai:1994kq}, which allow the Dirac operator to extend from $\IS^1$ to $\ID^2_+$. This is the same as requiring that the fermions are extended to the 6d manifold with boundary. By the usual arguments\footnote{A quick recap for the unfamiliar reader: the asymptotic holonomy that a spinor picks up in a cigar due to the curvature of the spin connection is  $1/2$ (because of the spin) of the integral of the curvature, namely the Euler characteristic of the disc $\chi(\ID^2)=1$. Hence, the spinor picks up a factor $e^{\frac 12 2\pi i}=-1$.} about the change in the periodicity between flat space and a cigar, this implies that the fermions are antiperiodic. 

Hence in the dimensional reduction, there are no zero modes on the KK reduction of the 5d matter fields, and hence no massless 4d half hypermultiplet in the bifundamental, and hence no 4d Witten global gauge anomaly. 

The ETW brane has managed to evade the 4d global gauge anomaly, in a fairly non-trivial way from the higher- dimensional perspective, which involves the interplay with the inflow that cancels the 5d global gauge anomaly. 
It is remarkable how the different pieces of information assemble in this intricate example to make things work.
On the other hand, the situation is straightforward from the perspective of the 4d theory: the massless spectrum is simply directly anomaly free. This is actually expected from the viewpoint of the 5d AdS$_5$ theory: the 5d bulk simply does not have non-trivial topological couplings to support any inflow, as is easily verified by the existence of very simple 4d boundaries which are automatically anomaly free. Hence, no matter how complicated an ETW configuration we manage to construct form the higher 10d dimensional perspective, when we reduce it to a 4d boundary of a 5d AdS$_5$ bulk it must necessarily come out as free of any anomaly, besides the mixed 1-form anomaly implicit from the coupling (\ref{bf-boundary}) coming from (\ref{bf-symtft}) in the 5d bulk. We will develop this more in section \ref{sec:lesson}.

\subsubsection{A remark on odd number of fermions for monopole operators}
\label{sec:odd-remark}

In this section we would like to emphasize an interesting spin-off of our above discussion, related to the proposed inconsistency of theories containing solitons with an odd number of fermions \cite{Sato:2022vii}. Our configuration of D3-branes ending on the fork configuration for $C_0=1/2$ provides an interesting loophole for this statement (or more precisely, for naive higher-dimensional extensions of it).

Let us recap the result in \cite{Sato:2022vii}. The statement is that a theory with 3+1 and higher dimensions is inconsistent if it contains dynamical point-like solitons with odd number of Majorana fermion zero modes. The general argument is that a topological obstruction to define a wavefunction for a system of two such solitons. A prototypical application of this inconsistency criterion is 4d $SU(2)$ theory with an odd number of fermions in the fundamental representation. Adding an scalar in the adjoint, one can build a gauge monopole which inherit an odd number of fermion zero modes from the flavours, so it is inconsistent according to the criterior. This of course fits nicely with the familiar inconsistency of this 4d theory due to Witten's global gauge anomaly from $\Pi_4(SU(2))=\IZ_2$ \cite{Witten:1982fp}. A similar example in 5d is $SU(2)$ with an odd number of fermions in the fundamental representation. In this case the point-like soliton is given by a 4d gauge instanton, which is a particle from the 5d perspective, and which inherits  an odd number of fermion zero modes from the flavours, so it is inconsistent according to the criterion. This of course fits nicely with the familiar inconsistency of this 5d theory due to a global gauge anomaly from $\Pi_5(SU(2))=\IZ_2$.

One may be tempted to naively extend the result by analogy and claim the inconsistency of a theory that contains any lower-dimensional defect with an odd number of worldvolume fermions. This would be problematic for our configuration of D3-branes ending on the fork for $C_0=1/2$, because the boundary of the D3-branes on the 5-branes may be regarded as a (Dirac) monopole of the 5d gauge theory \cite{Hanany:1996ie} with an extended 3d worldvolume, with and odd number of fermion zero modes inherited from the flavours. Actually, the resolution of this puzzle is that the analogue of the non-trivial topological obstruction in the soliton system is played by the global gauge anomaly in the 3d theory, which is in turned solved by the introduction of the 3d Chern-Simons coupling. Hence, the additional dynamics arising from worldvolume gauge fields makes the monopole consistent. 

Incidentally, the mechanism is reminiscent of a phenomenon familiar in the string theory realization of topological operators for non-invertible symmetries, where the definition of the operators involves non-trivial wordvolume dynamics of the branes realizing the topological operator \cite{Apruzzi:2022rei,GarciaEtxebarria:2022vzq}. It would be interesting to explore this relation further.

\subsection{Gravity dual and SymTFT of the 3d parity anomaly}
\label{sec:parity-anomaly}

The dynamical fields on the ETW brane are just the $USp(2n_+)\times SO(2n_-+1)$ gauge group, with no massless matter coming from the 5d multiplets. By extension, we expect a similar result for the case $C_0=0$, with just the $USp(2n_+)\times SO(2n_-)$ gauge group. Recall from section \ref{sec:fork-bc-parity-anomaly} that a genuine difference between the two cases
is the appearance of the 3d parity anomaly, which in field theory arises from the effect of the flavours in the D3-D5' open string sector. In the holographic dual the D3-branes have been geometrized, and their quantum effects should arise from supergravity. In the following we argue that it is easy to recover couplings such as (\ref{parity-D5}) from the gravity dual picture. Since this involves purely topological couplings, it is part of 4d SymTFT of the 3d BCFT$_3$ located at the ETW boundary of the 5d SymTFT of the complete system, in a nested picture in the spirit of \cite{Cvetic:2024dzu}. 

Consider the D5'$_-$-branes on top of the O5'$^-$-plane, which are wrapped on AdS$_4\times\ID^2_-$, leading to a 4d $SO(2n_-+1)$ gauge symmetry. Let us regard the configuration in the double cover, for simplicity, so we regard the D5'-branes split in $\IZ_2$ images (plus one unpaired). As usual, the gauge coupling is related to the inverse of the wrapped volume, and there is also a non-trivial $\theta$ angle. The latter arises from the reduction of the D5'-brane worldvolume topological 6d term, which for each half of the $\IZ_2$ symmetric set reads
\beqa
\int_{6d} C_0 {\cal F} \tr F^2\quad \Longrightarrow\quad \frac 12 n\int_{4d} \tr F^2\, .
\label{theta}
\eeqa
Here the trace is regarded in the unitary group of the half set of D5'-branes. In the second setup we have replaced $C_0=1/2$, and noticed that the integral of ${\cal F}$ over $\IS^2$ measures the number $n$ of D3-branes ending on the D5'-branes (with the $n$ image D3-branes ending on the image D5'-branes). Combining the result from the two $\IZ_2$ image sets we get a $\theta$ angle as in (\ref{theta}), now interpreted in $SO(2n_-+1)$. This 4d topological term in AdS$_4$ precisely reproduces the Chern-Simons term (\ref{parity-D5}) in the holographic 3d boundary, hence reproducing the part of the 3d parity anomaly manifest on the fields in the gravity dual.

As a final comment, let us remark that this analysis illustrates that the 4d ETW boundary can support topological couplings besides (\ref{bf-boundary}), as long as they involve fields that propagate only on the ETW brane boundary. In other words, if the same bulk theory has other boundaries, such terms need not arise on the additional boundaries, because they are model-dependent (i.e. depending on the microscopic details of the boundary), and not simply inherited from the bulk.

\section{General lessons: Topological Constraints on ETW branes}
\label{sec:lesson}

In this section we draw a general lesson learnt from the above system. It is based on a simple relation between the topological sectors of bulk and boundary theories, which can be read as a powerful constraint on the possible ETW boundaries in a given bulk theory.

\subsection{The general argument}
\label{sec:argument}

In section \ref{sec:puzzle-resolved} we made the key observation that there is a fairly intricate ETW configuration for the 10d AdS$_5\times\IS^5$ asymptotic region, involving non-trivial cancellation of localized contributions to a 5d global gauge anomaly via inflow, which however simplifies enormously when reduced to a 4d boundary of the 5d AdS$_5$ bulk theory. As explained there, the underlying reason for this drastic simplification is that there is a very limited set of topological couplings in the 5d bulk theory, and this limits drastically the structure of anomalies that a 4d boundary can support. In fact, following the same arguments, such structure must be universal, i.e. independent of the microscopic details, for different ETW configurations of the 4d theory as long as they provide consistent 4d boundaries for the theory. At the topological level, the only freedom in the physics of the ETW boundary therefore shows up in the presence of topological couplings involving only fields localized on the 4d boundary itself, such as the $\theta$ term for the $SO(2n_-+1)$ theory related to the 3d parity violation.

We can turn the above argument around to obtain a powerful constraint on allowed ETW configurations in a given theory.
In a general $d$ dimensional theory, we may be interested in classifying all possible $(d-1)$-dimensional ETW configurations and understanding their dynamics. This general question can be addressed in theories not coupled to gravity, but, in the light of the Swampland Cobordism conjecture \cite{McNamara:2019rup}, it is most meaningful in the context of theories of quantum gravity, where such ETW configurations are conjectured to exist. 

This completely general characterization may in general be a too ambitious goal, so we can restrict it to the topological sector. At this point, our argument above provides an interesting restriction, as it fixes the structure of anomalies on any $(d-1)$-dimensional ETW boundary in terms of the universal topological inflows allowed by the $d$ dimensional bulk theory (this of course includes the possibility that there is absolutely no inflow, in which case any ETW boundary must be automatically free of all anomalies). Hence, this provides an interesting non-trivial constraint on the microscopic realization of ETW boundaries, which simplifies the classification problem. 

Moreover, our arguments above show that, once this universal sector is fixed, the freedom to endow the ETW boundaries with additional topological couplings is restricted to terms involving only fields localized on the ETW boundaries themselves. Both this localized field content and the corresponding topological terms can vary among the different kinds of ETW branes in a given theory, as long as they are compatible with the universality of inflows.

In the next section we revisit the different known kinds of ETW branes for AdS$_5\times\IS^5$, and their interplay as 4d ETW boundaries, paying special attention to their topological couplings to the 5d bulk.

\subsection{Some explicit examples}
\label{sec:further}

In this section we would like to provide some concrete examples of gravity theories admitting several consistent ETW branes, which must hence satisfy the above criteria. To make this even more transparent, we build examples in which the different ETW branes are present simultaneously, namely, we consider a bulk theory with several boundary components. Finally, we consider our bulk theory to be given by AdS$_5(\times\IS^5)$, which has been the main arena for our previous discussions. 

As we have reviewed, there are several classes of BCFT$_3$ boundary conditions for which the gravitational dual has been described: (A) The duals of 3d $\NN=4$ configurations with NS5- and D5-branes reviewed in section \ref{sec:d3-ns5-d5-dual}; (B) the duals of 3d $\NN=4$ configurations with NS5-, D5-branes, with an O5-plane reviewed in section \ref{sec:o5-dual}; (C) the duals of 3d $\NN=2$ configurations with NS5-, D5- with O5'/D5' forks, discussed in section \ref{sec:d3-ns5-d5-o5-dual}. This is an obviously incomplete list of all the possible boundaries that the bulk AdS$_5(\times\IS^5)$ admits (in particular, there is no known gravity dual of the boundary conditions of NS5-, D5-, NS5'- and D5'-brane configurations in section \ref{sec:rotated-ns5-d5}), but is suffices to illustrate the key points.

\begin{figure}[htb]
\begin{center}
\includegraphics[scale=.08]{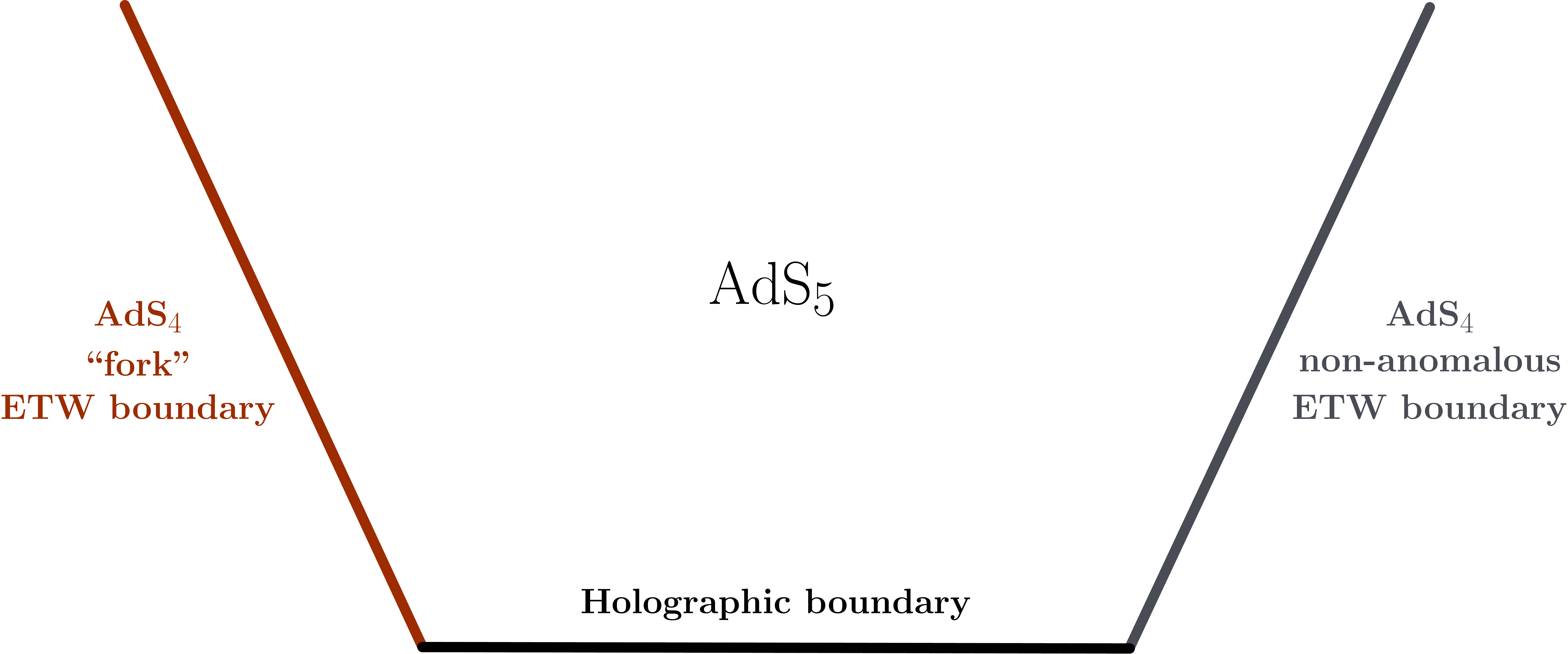}
\caption{\small AdS$_5$ with two ETW AdS$_4$ boundaries. }
\label{fig:wedge}
\end{center}
\end{figure}

As we have emphasized, the ETW branes should have topological couplings consistent with the fact that they are all boundary theories of the same bulk theory. One way to make this manifest is to build solutions in which several of these boundaries coexist. One simple mechanism to introduce several boundaries in our setup, which allows for a simple holographic dual, is the wedge holography \cite{Akal:2020wfl,VanRaamsdonk:2021duo}, briefly mentioned in section \ref{sec:puzzle-d5}. Namely, we consider D3-branes stretched between two different sets of 5-branes (possibly with O5-planes) leading to two different boundary conditions. For fairly general choices of the parameters of the boundary configurations, the holographic dual is a large wedge (in Poincar\'e coordinates) of AdS$_5(\times\IS^5)$ bounded by the two AdS$_4$ ETW branes dual to the BCFT$_3$'s given by the 5-brane configurations, see Figure \ref{fig:wedge}. In the following we provide several classes of examples of this kind:

One possibility is to consider the two ETW branes to be of type (A). Actually this situation has already been considered in detail in e.g. \cite{VanRaamsdonk:2021duo,Karch:2022rvr}, leading to large classes of AdS$_5(\times\IS^5)$ wedges bounded by two AdS$_4$ ETW branes dual to the Gaiotto-Witten BCFT$_3$.

We can consider setups in which the direction 3 along which the D3-branes stretch is made compact, and is acted on by an orientifold introducing two O5-planes sitting on opposite sides of the $\IS^1$. 
The compactification turns the brane configurations into an elliptic model \cite{Witten:1997sc}, which have been explored in setups preserving 8 supersymmetries \cite{Brunner:1997gf,Hanany:1997gh,Uranga:1998uj,Park:1998zh} and 4 supersymmetries \cite{Park:1998zh} (see also our discussion in Appendix \ref{sec:tduals}). The former can be regarded as related to gravity duals with the two ETW branes of type (B), while the latter have both ETW branes of type (C).

In the following, we would like to mention that it is possible to have mixed situations, and build a concrete class of models in which one of the ETW branes is of type (C) while the second is of type (A). We will consider the first to correspond to an O5'/D5' fork. Let us now discuss the second ETW boundary, which we describe in the covering space of the orientifold imposed by the first. Namely, we choose a configuration of 5-branes at $x^3=L$, and an orientifold image configuration at $x^3=-L$. Note that, since we want to stretch D3-branes from the O5'/D5' fork to these 5-branes, one should be careful to satisfy the s-rule, which states that between 5-branes of different type, say $(p,q)$ and a $(p',q')$, there can be at most $|pq'-p'q|$ suspended D3-branes. The simplest way to avoid a possible clash with the s-rule is to have the second ETW brane to involve at laeast some branes of the same kind as those in the first ETW brane. 

Let us consider the simplest case of $C_0=0$ first, and choose the simple fork configuration in which there are 4 D5'$_+$-branes, so that charges are trivially conserved in the fork, and the 5-brane web is a trivial intersection of an NS5-brane with the O5'/D5' system. In order to build the second ETW brane, we can then locate stacks of NS5- and D5-branes at (slightly separated positions around) $x^3=L$ and their orientifold images at $x^3=-L$, leading to a local ETW brane of type (A) in the quotient. Treating the O5'-plane and the D5'-branes as probes, one can build an explicit supergravity solutions for these configurations, realizing the wedge holography setup of AdS$_5(\times\IS^5)$ with two AdS$_4$ ETW boundaries.

\begin{figure}[htb]
\begin{center}
\includegraphics[scale=.09]{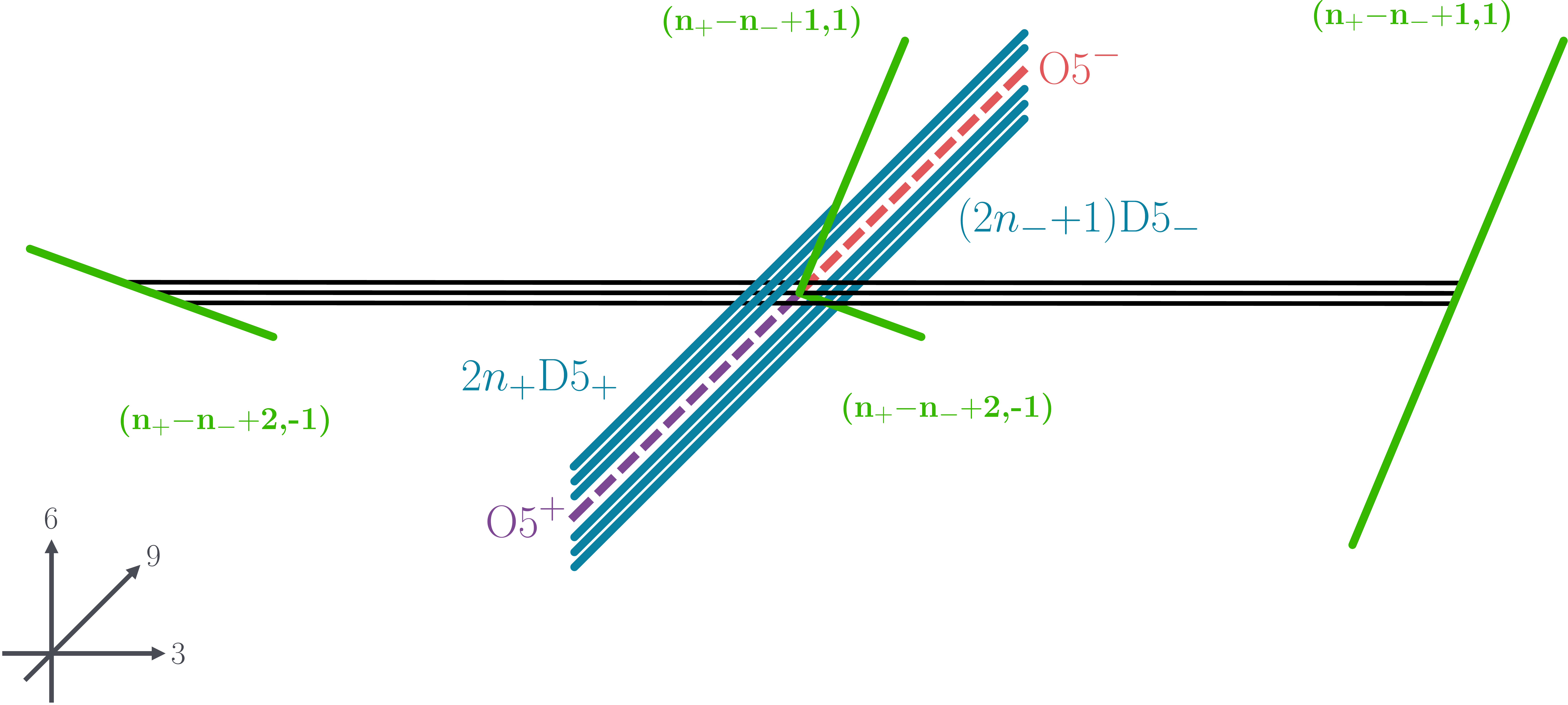}
\caption{\small A simple example of brane configuration dual to AdS$_5$ with two ETW AdS$_4$ boundaries, one given by the anomalous fork configuration, a second one given by some non-anomalous boundary configuration. }
\label{fig:two-ads4-etws}
\end{center}
\end{figure}

We can also consider the case $C_0=1/2$, which is more interesting. In a general configuration for the ETW brane at $x^3=0$ there is one half $(p,1)$ 5-brane and its orientifold image half $(p+1,-1)$ 5-brane (they are basically the original NS5-brane, suitably modified to absorb the RR 5-brane charges of the O5'$^{\pm}$-planes and the D5'$_\pm$-branes, recall Figure \ref{fig:fork-general-12}, with $p=n_+-n_-+1$).  In order to build the second ETW brane, we can then locate one (whole)  $(p,1)$ 5-brane at $x^3=L$ and its orientifold image (whole) $(p+1,-1)$ 5-brane at $x^3=-L$. In addition we may add extra $(p,1)$ 5-branes and D5-branes (really meaning unprimed ones) at $x^3=L$, leading to a local ETW brane of type (A) (albeit in a different $SL(2,\IZ)$ frame), and their orientifold image $(p+1,-1)$ 5-branes and D5-branes at $x^3=-L$. At weak coupling, one can build an explicit supergravity solutions, since the $(p,1)$ and $(p+1,-1)$ 5-brane behave as NS5-branes.
The configuration is shown in Figure \ref{fig:two-ads4-etws}.

We hope these examples suffice to illustrate the construction of bulk theories with several boundaries. These constructions make the analysis in section \ref{sec:puzzle-d5} more explicit, and make manifest the fact that different ETW branes of the same bulk theory must have the same structure of topological inflows. Although it is clearly possible to carry out those arguments in other gravitational backgrounds which are not necessarily related to AdS or to holography, in this work we have chosen to stick to the holographic setup, leaving a more general exploration for future work.

\section*{Acknowledgments}

We are pleased to thank I\~naki Garcia-Etxebarria and Miguel Montero for many useful discussion and collaboration in closely related topics. We also thank Roberta Angius, Matilda Delgado, Luis Ib\'a\~nez, Fernando Marchesano, Juan Pedraza, Irene Valenzuela, Chuying Wang and Xingyang Yu for useful discussions.  
This work is supported through the grants CEX2020-001007-S and PID2021-123017NB-I00, funded by MCIN/AEI/10.13039/501100011033 and by ERDF A way of making Europe. The work by J. H. is supported by the FPU grant FPU20/01495 from the Spanish Ministry of Education and Universities. 

\newpage 

\appendix 

\section{Global anomaly inflow in 4d brane configurations}
\label{sec:4d-global-anomaly}

The class of fork configurations in section \ref{sec:fork-c0} has analogues in the case of configurations of D4-branes ending on NS5-branes, in the presence of O4-planes. This setup also illustrates the appearance of localized contributions to a global gauge anomaly, in an even simpler setup, since the backreaction of D4-branes (and O4-planes) on NS5-branes is logarithmic \cite{Witten:1997sc}, so no non-trivial brane webs are involved.

The setup is obtained by formally T-dualizing the 5-brane configurations in one of the 5d Poincar\'e invariant directions. For convenience we switch to more standard conventions following \cite{Witten:1997sc}. Hence we consider NS5-branes along the directions 012345 with suspended D4-branes along 0123 and finite extent in 6, and we also introduce O4-planes along 01236. There is Poincar\'e invariant in the 4d space 0123, the NS5 branes have two dimensions (45) transverse to the D4-brane boundaries, so brane bending is only logarithmic and the NS5-brane stays an NS5-brane in the whole configuration. The O4-plane flips its RR charge as it crosses an NS5-brane \cite{Evans:1997hk}. The analogue of the fork configuration is thus obtained by taking one NS5-brane and one O4-plane flipping sign across it. We consider  $2n_+$ semi-infinite D4$_+$-branes on top of the semi-infinite O4$^+$-plane, and $m_-$ D4$_-$-branes on top of the O4$^-$, with $m_-$ even or odd. The gauge symmetry is $USp(2n_+)\times SO(m_-)$, and there is a bifundamental half-hypermultiplet at the intersection. There are $m_-$ flavours in the fundamental of the $USp(2n_+)$, so for odd $m_-=2n_-+1$ there is a local contribution to the 4d Witten's $\IZ_2$ global gauge anomaly associated to $\Pi_4(USp(2n))=\IZ_2$ \cite{Witten:1982fp}. By the analogue of the arguments in the 5-brane case, there should be an inflow\footnote{We are grateful to I\~naki Garcia-Etxebarria and Miguel Montero for very useful discussions on  global gauge anomaly inflows.} of global anomaly from the 5d$_+$ bulk. In this case, the 5d anomaly theory is the $\eta$ invariant \cite{Witten:1982fp}, so we expect that on the volume of the D4$_+$/O4$^+$ system, we have a topological term
\beqa
\frac 12 \int_{5d_+} \eta_5\, .
\eeqa
This is to our knowledge the first string theory realization of global anomaly inflow \cite{Witten:2019bou} for the 4d $\IZ_2$ global gauge anomaly of \cite{Witten:1982fp}. The above argument is an interesting way to complete, in the spirit of the inflow arguments of \cite{Green:1996dd}, the topological couplings of D-branes and O-planes. On the other hand, it would be interesting to derive the presence of the above coupling from first principles, using a suitably K-theory interpretation the topological couplings on D-branes \cite{Minasian:1997mm,Witten:1998cd,Moore:1999gb} and O-planes \cite{Freed:2000ta}.

\section{T-duals of O5/D5 fork configurations}
\label{sec:tduals}

The configuration of the fork for $C_0=1/2$ may result somewhat unfamiliar. In this appendix we relate it to other systems via circle compactifications and T-duality, in order to put it in a wider context, and to illustrate how its properties nicely fit with non-trivial consistency conditions of higher-dimensional branes considered in the literature.

\subsection{Recap of the setup}

We start with a recap of the setup and conventions. For convenience to match T-dual conventions, in this appendix we drop the primes for D5-branes and O5-planes. We consider type IIB with a $(p,q)$ 5-brane configuration, including O5-planes, which flip sign when split in halves by a $(p,1)$ 5-brane (i.e. 1 unit of NS5-brane charge). The directions are: the 5-branes and O5-planes span the directions 012 45 and one direction in the 2-plane 69. Using $z=x^9+i x^6$ the direction for a $(p,q)$ 5-brane is $z=p+\tau q$, with $\tau=C_0+i/g_s$. Because of the orientifolds (under which $C_0$ is odd) we will mainly focus in the cases of $C_0=0$ (so D5-branes and O5-planes run along 9 and NS5-branes run along 6) and $C_0=1/2$ (so D5-branes and O5-planes run along 9 and NS5-branes run in a direction close to 6 but tilted in the direction 9).

We are interested in fork configurations, in which the O5-plane is split by an NS5-brane (possibly bound to some D5-brane charge) in two halves denoted by O5$^\pm$, according to the sign of their RR charge. We also include in general $n_\pm$ $D5^\pm$-branes, i.e. half D5-branes on top of those O5$^\pm$-plane pieces.

 For the  fork at $C_0=0$ the number of D5$^\pm$-branes is even, while for the fork at  $C_0=1/2$ the number of D5$^-$-branes is odd. As mentioned in the main text, this is related to the fact that for $C_0=1/2$ the O5-planes actually correspond to the tilded version, although we momentarily abuse language and omit it. The number of D5$^+$ branes is even in any of these cases, since its worldvolume gauge group is $USp(2n_+)$. For $C_0=1/2$ there is a localized contribution to the 5d $\IZ_2$ global gauge anomaly for this $USp(2n_+)$, because it has an odd number of flavours, but we assume it is cancelled by a global anomaly inflow as explained in the main text, and we will not worry about it any more.
 
 Eventually we will be interested in discussing the introduction of D3-branes, which will span 012 and will be semi-infinite in 3. Actually the orientifold requires to have such semi-infinite D3-branes both at $x^3>0$ and their orientifold images at $x^3<0$. If necessary, these two sets of semi-infinite D3-branes can be recombined and moved off along some of the 5-branes. This can be made even more dynamically if the D3-branes are made of finite extent in 3 by locating additional 5-branes at $x^3\neq 0$.
 
\subsection{T-duality along 6}
\label{sec:t6}

We start by considering the fork for O6/D6 systems and relate them to via T-duality to forks for O5/D5 systems. We will find a precise match with the consistency conditions for O6-planes in \cite{Hyakutake:2000mr}.

Consider IIA theory in flat 10d and consider one NS5 brane along 012 456 stuck at an O6-plane spanning 012 456 9. Note that the NS5-brane splits the O6-plane into two O6$^\pm$ halves stretching along $x^9\lessgtr 0$. We introduce $n_{D6_\pm}$ D6$_\pm$-branes (as counted in the double cover) on top of the half O6$^\pm$. In pure type IIA theory we should have $n_{D6^-}=n_{D6^+}+8$ to ensure conservation of RR charge, but in the presence of a non-trivial Romans mass $m$ the condition is changed to $n_{D6^-}=n_{D6^+}+8-m$ \cite{Hanany:1997sa}.

Note that for even (resp. odd) $m$ we have an even (resp. odd) number of D6-branes on top of the O6$^-$. This is in agreement with the result in \cite{Hyakutake:2000mr} that the ${\widetilde{O6}}^-$-plane (O6$^-$ with one stuck D6-brane) is possible only if $m$ is odd, and the O6$^-$ with no stuck D6-brane is possible only if $m$ is even. There is a similar statement for O6$^+$ vs ${\widetilde{O6}}^+$, which is also satisfied in these fork configurations.

We also note that on top of the NS5, the fork introduces a 6d $\NN=1$ half-hypermultiplet in the bifundamental of the $USp(n_{D6_+})\times SO(n_{D6_-})$, so there is an odd number of 6d flavours for the $USp(n_{D6_+})$. However, in 6d there is no global gauge anomaly for these groups, so there is actually no global anomaly inflow in this setup\footnote{Even for $USp(2)=SU(2)$, although $\Pi_6(SU(2))=\IZ_{12}$ \cite{Bershadsky:1997sb}, there is no actual anomaly, because $\Omega_7^{Spin}(BSU(2))=0$, see \cite{Lee:2020ewl} for discussion.}. 

We now consider these O6/D6 fork configurations, compactify them along the direction 6 (i.e. along a common direction of the NS5-brane and the O6-plane) and perform T-duality to obtain a configuration with O5/D5 forks. We note that this kind of T-duality for the O6/D6 fork has been considered in the context of brane constructions for 6d SCFTs in \cite{Hayashi:2015vhy}. For simplicity we focus on $m=0,1$, in which case the number of D6$_-$-branes is 8 (resp. 7) units bigger than the number of D6$_+$-branes. 

Upon compactification of the direction 9 and T-duality, the NS5-brane remains an NS5-brane, the O6-planes turn into two O5-planes at opposite points in the T-dual circle, and the half D6-branes turn into half D5-branes which are located at either of the two O5-planes, and which we choose to distribute as equitatively as possible, for simplicity. So the D6-brane fork turns into two 5-brane forks.

\begin{figure}[htb]
\begin{center}
\includegraphics[scale=.069]{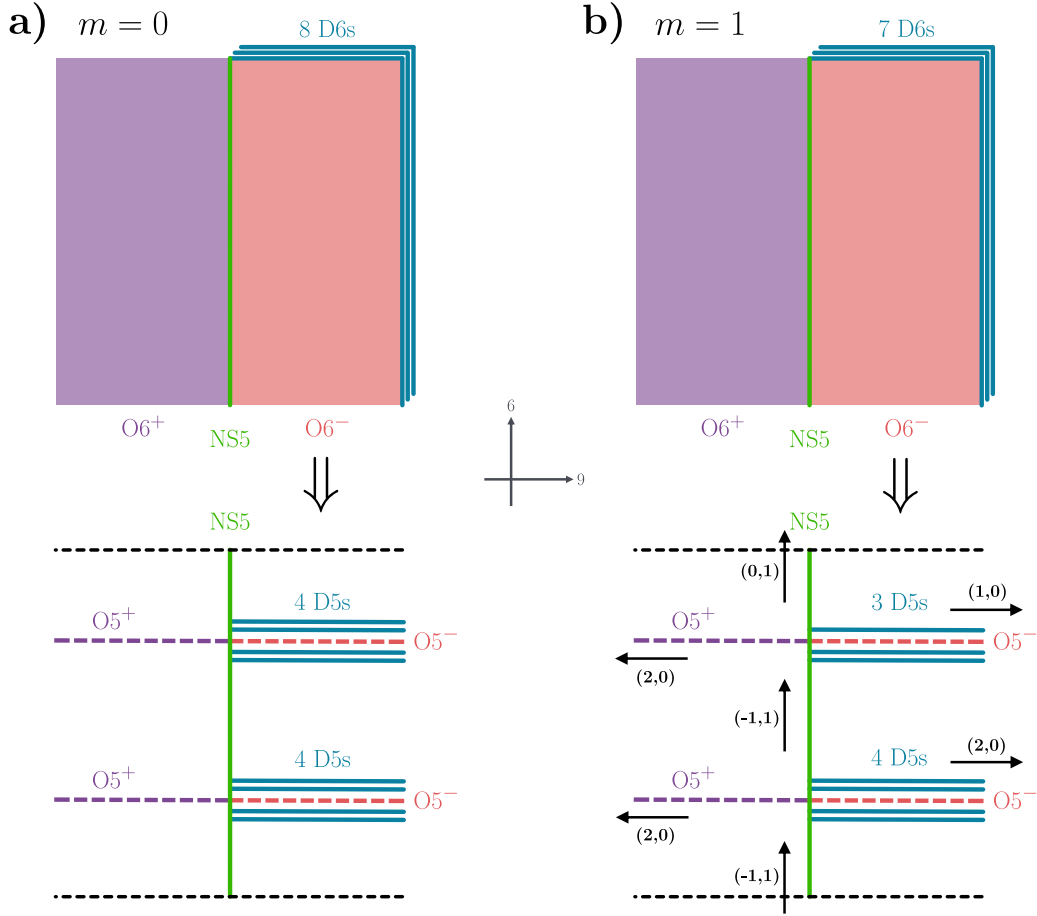}
\caption{\small O6/D6 fork and its T-duality to O5/D5 forks (a) for $m=0$ and (b) for $m=1$ (in the latter, for simplicity we have ignored brane bending, which is a good depiction for small $g_s$). In the lower pictures the thin black dashed lines are identified to make the vertical direction a circle. Note that in case (b) there must be a non-trivial holonomy $C_0\to C_0+1$ on the circle so that the bottom $(-1,1)$ brane can be identified with the top $(0,1)$-brane. The fact that $\int_{\IS^1} dC_0=1$ matches the value $m=1$ in the original picture.}
\label{fig:t6}
\end{center}
\end{figure}

We start with the simpler case of $m=0$, see Figure \ref{fig:t6}a. In this simplest configuration, the O6$^+$/O6$^-$ fork splits into two O5$^+$/O5$^-$ forks at $C_0=0$, and 8 D6$_-$-branes split into 4 D5$_-$-branes on each such fork. Hence the D5-brane charges  match on both sides of the NS5-brane, and the $(p,q)$ 5-brane web is trivial. 
 
For $m=1$ the story is similar but slightly trickier, see  Figure \ref{fig:t6}b. There are only 7 D6$_-$-branes, so in the T-dual 4 of them can go to one O5/D5 fork, to produce one trivial 5-brane web corresponding to a $C_0=0$ fork, but the 3 remaining ones make the second fork unbalanced, so that it corresponds to a $C_0=1/2$ fork. This however matches perfectly the fact that the T-dual of the Romans mass $m=1$ is a non-trivial flux $\int_{\IS^1} dC_0=1$ in the T-dual circle, which then implies that there must indeed exist one fork with $C_0=0$ and one with $C_0=1/2$.

Incidentally, we may come back to the case of $m=0$ and realize that it is possible to T-dualize it to two forks at $C_0=1/2$ (which is compatible with $m=0$ in the initial model) if we allow for unequal distribution of the T-duals of the 8 D6-branes, by locating e.g. 3 in one of the O5/D5 forks and 5 in the second. Which of the D-brane distributions arises is fixed by the choice of D6-brane Wilson lines on the $\IS^1$ upon compactification, so we stick to the simplest choice of equitative distributions.

We also note that the above T-dualities can be carried out in the presence of extra branes turning into D3-branes in the type IIB side, namely D4-branes along 01236. Such brane configuration are related those realizing 4d $\NN=1$ theories with chiral spectrum from the fork \cite{Landsteiner:1998gh,Brunner:1998jr,Elitzur:1998ju}. At this point, we may with to take the limit of large circle and decouple one of the 5-brane forks, so as to reach the configurations considered in the main text. 

 \subsection{T-duality along 3}
 \label{sec:t3}
 
 In this section we start with the configuration with O5/D5 fork and make the direction 3 compact. This may be of interest as in section \ref{sec:further} to study setups of wedge holography with two ETW branes. In this appendix we emphasize other motivation, which is applying T-duality along the direction 3 to relate the O5/D5 forks to other setups in the literature. 
 
 Because the O5-plane flips the direction 3, there is another orientifold fixed point when it is made compact, leading to different possible compact models, which we discuss in the following. We note that a similar discussion of brane configurations and T-dualities was studied in \cite{Park:1999eb}, in a system of O6/D6 forks with D4-branes along the compact direction (related by a formal T-duality in 9), so we can basically translate those results. 
 
 The basic ingredients in the T-duality are:
 \smallskip
 
 $\bullet$ The NS5-branes T-dualize into Taub-NUT geometries, so if we have a total of $N$ NS5-branes (as counted in the double cover), the T-dual will be a $\IC^2/\IZ_N$ geometry, where the generator $\theta\in\IZ_N$ acts as $\theta:(z_1,z_2)\to(e^{2\pi i/N}z_1,e^{-2\pi i/N}z_2)$. Alternatively we can describe the geometry by using invariant coordinates $u=z_1^N$, $v=z_2^N$, $z=z_1z_2$, with $uv=z^N$.
 
 \smallskip
 
 $\bullet$ The O5-planes T-dualize, roughly speaking, to O6-planes. In general the orientifold action flips the coordinate $z\to -z$ and flips one extra coordinate corresponding to 9. The O-plane sits at $z=0$, namely $uv=0$, so it splits into two components parametrized by $u$ (with $v=0$) and by $v$ (with $u=0$), reproducing the T-dual of the fork splitting. The D5$^\pm$-branes will be dual to D6-branes on either of these 2-cycles as well. The different possibilities of O5-planes and D5-branes will be discussed shortly. 
 
 One novelty is that we can discuss configurations with equal or different values of $C_0$ at the two O5/D5 fork. If the value of $C_0$ is the same for the two O5-planes (both $C_0=0$ or both $C_0=1/2$), then we can take $C_0$ constant, and this simply turns into a constant background for the RR 1-form in the IIA T-dual. On the other hand, if $C_0=0$ on top of one O5-plane and $C_0=1/2$ on top of the other O5-plane, then there is a non-trivial $dC_0$ along the (double covering) $\IS^1$, with $\int_{\IS^1}dC_0=1$; this means that in the T-dual there is 1 unit of RR 0-form flux $m$ (Romans mass). This will be crucial to  match the fact that an O6$^-$-plane admits an odd (respectively even) number of D6-branes on top if $m$ is odd (respectively even) \cite{Hyakutake:2000mr}.

We now consider several classes of models in turn. In the comparison with \cite{Park:1999eb}, we restrict to examples which include at least one O5/D5 fork.

\subsubsection{Odd number of NS5-branes}
\label{sec:t3-odd}

Consider the case of an odd number of NS5-branes in the direction 3. Because of the orientifold action, one of the NS5-branes is on top of an O5-plane, producing a fork, while the other O5-plane has no NS5-brane and hence no fork. The latter O5-plane may be of positive or negative charge). We note that there is one half-direction in 9 where the two O5-planes at the opposite sides of the $\IS^1$ have the same sign, while in the other half their signs are opposite. This will have a reflection in the T-dual.

The T-dual is an orientifold of $\IC^2/\IZ_N$ with orientifold group
\beqa
(1+\theta+\ldots+\theta^{N-1})(1+\Omega (-1)^{F_L}{\cal R}_1 R)\, ,
\eeqa
where ${\cal R}_1:z_1\to -z_1$ and $R$ flips 6. The orientifold action has a fixed plane parametrized by $z_2$ (equivalently, $v$, at $u=0$), so there is an O6-plane with positive or negative charge there; this is the T-dual of the half direction where the two O5-planes had the same charge. On the other hand, there is no orientifold plane parametrized by $u$ at $v=0$; this is because in the other half direction  the O5-planes have opposite charges that hence cancel, so  explicit O6-plane arises in the T-dual.

The relation of the $C_0$ values at the O5/D5 forks and the Romans mass $m$ in the T-dual is as follows:

$\bullet$ If both O5-planes have the same $C_0$ in both O5-planes (so we must have $m=0$ in the T-dual), then any O5$^-$ (either whole or half) must have an even (respectively odd) number of D5-branes on top if $C_0=0$ (respectively $C_0=1/2$). Upon T-duality, we have to combine the half O5-planes in the opposite ends of the $\IS^1$, and there are the three following possibilities: i) both are O5$^-$, in which case they T-dualize to an O6$^-$, with an even number of D6 on top, which is consistent with $m=0$; ii) both are O5$^+$, in which case they T-dualize to an O6$^+$, with an even number of D6, again compatible with $m=0$; iii) one O5$^+$ with one O5$^-$, in which case they T-dualize to no O6-plane, so even though we may get an odd number of D6-branes there is no contradiction with having $m=0$.

$\bullet$ If we have different $C_0$ in the two O5-planes (namely $m=1$ in the T-dual), then any O5$^-$ piece on top of $C_0=1/2$ will have an odd number of D5-branes on top, while there will be an even number of D5's on top of any O5$^-$ with $C_0=0$ or of O5$^+$-planes with any $C_0$. If we denote by an ordered pair $(\pm,\pm)$ the signs of the two O5-planes at $C_0=0$ (first entry) and $C_0=1/2$ (second entry), we have four cases: i) $(+,+)$, which T-dualizes to O6$^+$ with even number of D6's; ii) $(-+)$, which T-dualizes to no O6-plane and an even number of D6's; iii) $(-,+)$, which T-dualizes to no O6-plane with an odd number of D6's; iv) $(-,-)$, which T-dualizes to and O6$^-$-plane with an odd number of D6-branes. All these situations, in a particularly non-trivial way in case iv), are compatible with $m=1$.

\subsubsection{Even number of NS5-branes}
\label{sec:t3-even}

Consider now the case of an even number of NS5-branes in the $\IS^1$. It is possible to have models with no NS5-branes on top of the O5-planes (i.e. no forks), but, as explained, this is not particularly interesting for our purposes. Therefore we focus on the case with one NS5-branes on top of each O5-plane. These two forks can be oriented parallel or antiparallel, according to whether the charges of the O5-plane halves at e.g. $x^9>0$ are of the same sign for the two forks, or or opposite signs for the two forks.

For parallel forks, the T-dual orientifold group is 
\beqa
(1+\theta+\ldots+\theta^{N-1})(1+\Omega (-1)^{F_L}{\cal R}_1 R)\, .
\eeqa
In this case, on top of the O6-plane introduced by $\Omega (-1)^{F_L}{\cal R}_1 R$ (which wrap $z_2$, namely $v$ at $u=0$), there is an O6-plane introduced by $\Omega (-1)^{F_L}{\cal R}_2 R$, where  ${\cal R}_2\equiv {\cal R}_1 \theta^{N/2}$ acts as $z_2\to -z_2$, and produces an orientifold which wraps $z_1$ (namely $u$ at $v=0$). The fact that we get two O6-planes is dual to the fact that the signs of the O5-planes match for two parallel forks, and add up to O6-plane charges in the T-dual.

On the other hand, for antiparallel forks, the T-dual orientifold group is
\beqa
(1+\theta+\ldots+\theta^{N-1})(1+\Omega (-1)^{F_L}{\cal R}_1 R \alpha)\, ,
\eeqa
with $\alpha^2=\theta$. This $\alpha$ plays the role of a half-shift along the T-dual $\IS^1$, and implies that there are no fixed planes under the orientifold action, so no O6-planes. This is dual to the fact that the signs of the O5-planes mismatch and cancel for antiparallel forks.

The discussion of the equal or unequal $C_0$'s in this case is again in terms of the halves of O5-planes, their sign and the value of $C_0$ on top of them. So the discussion is identical to that in the previous section of odd $N$. Again we recover agreement with the rule in \cite{Hyakutake:2000mr} that O6$^-$ has an odd number of D6-branes on top if $m=1$, and an even number if $m=0$. 

\medskip

Either for even or odd $N$, one can now consider introducing D3-branes along 0123 in the original O5/D5 fork configuration, namely suspended between forks and follow the T-duality. The T-dual of this are D2-branes located at the corresponding orientifold quotients of $\IC^2/\IZ_N$. The corresponding gauge theories can be obtained as in \cite{Park:1999eb} and we skip further discussions here.

\bibliographystyle{JHEP}
\bibliography{mybib}

\end{document}